\batchmode
\documentclass[9pt,american]{extarticle}
\usepackage[LGR,T1]{fontenc}
\usepackage[utf8]{inputenc}
\usepackage{amsmath}
\usepackage{amssymb}
\usepackage[a4paper]{geometry}
\usepackage{array}
\usepackage{multirow}
\usepackage{booktabs}
\usepackage[colorlinks=true]{hyperref}
\makeatletter

\DeclareRobustCommand{\greektext}{%
  \fontencoding{LGR}\selectfont\def\encodingdefault{LGR}}
\DeclareRobustCommand{\textgreek}[1]{\leavevmode{\greektext #1}}

\usepackage{graphicx}
\usepackage{multicol}
\usepackage{authblk}
\usepackage{tikz-cd}

\author[1,2,*]{Zoltán Sántha}
\author[1,$\dagger$]{Gergő Thiering}

\affil[1]{\small HUN-REN Wigner Research Centre for Physics, PO Box 49, H-1525 Budapest, Hungary}
\affil[2]{\small Department of Atomic Physics, Budapest University of Technology and Economics,}
\affil[ ]{\small Budafoki út 8, H-1111 Budapest, Hungary}
\affil[*]{\small \textit{santha.zoltan@wigner.hun-ren.hu}}
\affil[$\dagger$]{\small \textit{thiering.gergo@wigner.hun-ren.hu}}

\makeatother

\usepackage{babel}
\begin{document}
\title{{\huge First-principles calculation of electron-phonon spectral functions
for defects}\\
{\huge using phonon interpolation}}
\date{\today}
\maketitle
\begin{abstract}
Accurate optical lineshapes of point defects require dense sampling
of lattice vibrations, which is challenging for direct first-principles
supercell calculations. We introduce a phonon-interpolation method
for calculating Huang-Rhys spectral densities and optical lineshapes
within the harmonic Franck-Condon framework. By exploiting the spatial
localization of transition-induced forces, the method obtains electron-phonon
coupling from a tractable defect supercell and samples its vibrational
response on dense reciprocal-space grids, avoiding explicit construction
and diagonalization of large hypercell dynamical matrices. Applied
to the negatively charged nitrogen-vacancy center in diamond, the
method resolves low-energy acoustic contributions, the dominant phonon sideband
structure, and finer optical-phonon features. The resulting temperature-dependent
photoluminescence lineshapes reproduce the main features of experimental
spectra after accounting for additional broadening. The approach offers
an efficient route to high-resolution defect spectra and can, in principle,
complement embedding and machine-learning methods.
\end{abstract}
\begin{multicols}{2}

\section*{Introduction}

Point defects in solid-state materials have attracted considerable
attention as promising platforms for quantum technologies, with potential
applications in quantum information processing \cite{wolfowiczQuantumGuidelinesSolidstate2021,taminiauUniversalControlError2014,childressDiamondNVCenters2013},
quantum sensing \cite{degenQuantumSensing2017,rondinMagnetometryNitrogenvacancyDefects2014},
quantum communication \cite{rufQuantumNetworksBased2021,bernienHeraldedEntanglementSolidstate2013,pompiliRealizationMultinodeQuantum2021},
and integrated quantum photonics \cite{sipahigilIntegratedDiamondNanophotonics2016,lukinIntegratedQuantumPhotonics2020,bogdanovMaterialPlatformsIntegrated2017}.
Their appeal arises from the possibility of combining atom-like electronic
transitions with the robustness and scalability of a crystalline host
\cite{wolfowiczQuantumGuidelinesSolidstate2021,awschalomQuantumTechnologiesOptically2018,atatureMaterialPlatformsSpinbased2018}.
In particular, optically active defects can enable optical initialization
and readout of quantum states \cite{dohertyNitrogenvacancyColourCentre2013,jelezkoObservationCoherentOscillations2004,rogersAllOpticalInitializationReadout2014},
as well as single-photon emission \cite{aharonovichSolidstateSinglephotonEmitters2016,kurtsieferStableSolidStateSource2000},
making their optical response a central criterion for evaluating their
technological relevance \cite{rufQuantumNetworksBased2021,awschalomQuantumTechnologiesOptically2018,atatureMaterialPlatformsSpinbased2018}.
Consequently, detailed characterization of defect-related optical
lineshapes is essential for understanding the underlying photophysical
mechanisms, including electron-phonon coupling \cite{alkauskasFirstprinciplesTheoryLuminescence2014,plakhotnikElectronphononProcessesNitrogenvacancy2015,jinPhotoluminescenceSpectraPoint2021},
spectral broadening \cite{jinPhotoluminescenceSpectraPoint2021,alkauskasFirstPrinciplesCalculationsLuminescence2012,mccullianQuantifyingSpectralDiffusion2022},
and interactions with the local solid-state environment \cite{mccullianQuantifyingSpectralDiffusion2022,rodgersMaterialsChallengesQuantum2021,orphal-kobinOpticallyCoherentNitrogenVacancy2023}.

A conventional theoretical route for interpreting defect-related optical
spectra is provided by the Huang-Rhys formalism, which expresses optical
transitions in terms of Franck-Condon overlaps between vibrational
states associated with different electronic configurations \cite{huangTheoryLightAbsorption1950,laxFranckCondonPrincipleIts1952}.
In this picture, the structural relaxation induced by an electronic transition
determines the coupling to lattice vibrations, while the resulting
phonon sideband reflects how the electronic transition is distributed
over the vibrational modes of the host \cite{zhangApplicationsHuangRhys2019,alkauskasTutorialDefectsSemiconductors2016}.
The formalism is therefore particularly useful because it connects
atomistic quantities, such as equilibrium geometries and normal modes,
to experimentally accessible observables including the zero-phonon
line intensity, the Debye-Waller factor, and the overall photoluminescence
lineshape \cite{alkauskasFirstprinciplesTheoryLuminescence2014,alkauskasFirstPrinciplesCalculationsLuminescence2012}.

For solid-state point defects, the microscopic input required for
this description is most commonly obtained from first-principles calculations,
typically within density functional theory combined with the supercell
approach \cite{freysoldtFirstprinciplesCalculationsPoint2014,vandewalleFirstprinciplesCalculationsDefects2004}.
In this framework, the defect is embedded in a finite periodically
repeated cell that approximates the dilute-defect limit while allowing
its local structural relaxation to be treated explicitly. However,
the finite size of the supercell also restricts the vibrational degrees
of freedom available to the calculation \cite{jinPhotoluminescenceSpectraPoint2021,razinkovasVibrationalVibronicStructure2021}.
Even relatively large supercells provide only a discrete sampling
of the host vibrational spectrum, with the longest-wavelength and
lowest-energy lattice vibrations either absent or represented only
incompletely.

This finite-size restriction is particularly important for optical
lineshape calculations because the phonon sideband is not determined
solely by strongly localized defect modes \cite{alkauskasFirstprinciplesTheoryLuminescence2014,silkinisOpticalLineshapesCenter2025}.
Extended host modes can also contribute to the spectral shape, especially
at low energies and in regions where the phonon density of states
exhibits pronounced structure \cite{silkinisOpticalLineshapesCenter2025,hashemiPhotoluminescenceLineShapes2021}.
Sparse vibrational sampling achievable directly from \textit{ab initio}
calculations may therefore obscure acoustic contributions, smooth
out or shift fine spectral features, and limit the extent to which
calculated lineshapes can be compared quantitatively or qualitatively
with high-resolution photoluminescence spectra \cite{jinPhotoluminescenceSpectraPoint2021,silkinisOpticalLineshapesCenter2025}.

\begin{figure*}[t!]
	\centering
	\includegraphics[width=\textwidth]{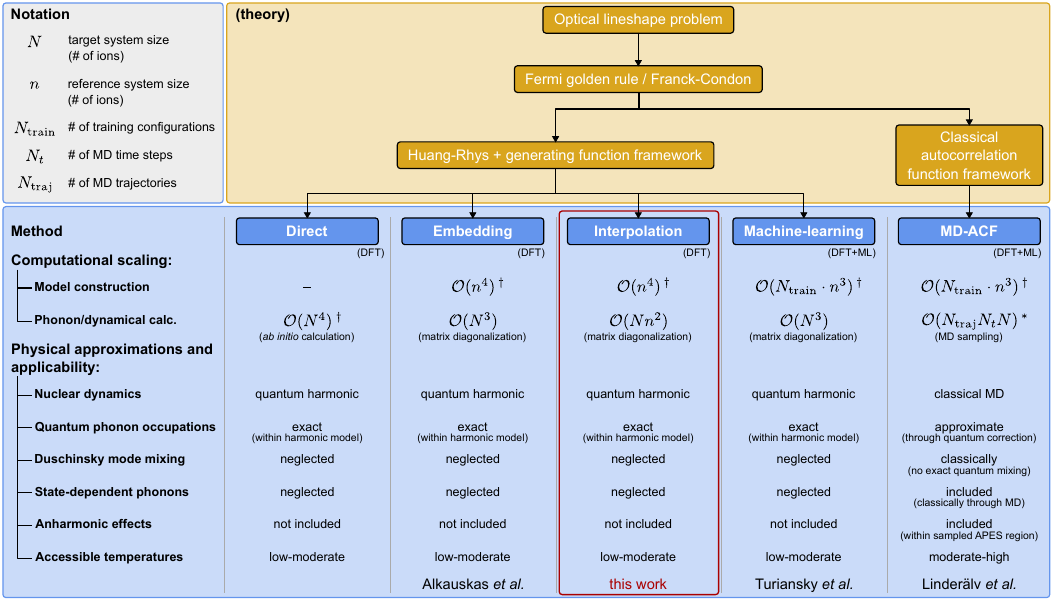}
	\caption{\label{figure1}\textbf{Relation of the present interpolation method to other approaches for optical lineshape calculations.} The methods are organized according to their theoretical framework and characterized by their computational scaling, relevant physical approximations and scope of applicability. \\
    $\,^\dagger$: Here an \textit{ab initio} DFT computational scaling of $\mathcal{O}(x^3)$ is assumed, where $x$ is the size of the simulated system. The appearing fourth order follows from applying DFT within the finite-differences scheme to obtain phonon modes.\\
    *: Achieving the same mean phonon-mode spacing across different methods requires $N_t\propto N$ for the MD-ACF method, yielding $\mathcal{O}(N^2)$ to $\mathcal{O}(N^3)$ scaling depending on whether $N_{\mathrm{traj}}$ remains constant or grows proportionally with $N$. \\
    (For further details on the computational scaling of the methods, see the Supplementary Information.)}
\end{figure*}

To obtain more detailed and converged optical spectra, several methods
have been introduced over the years. A prominent embedding approach was introduced
by Alkauskas \textit{et~al.} \cite{alkauskasFirstprinciplesTheoryLuminescence2014}
for investigating the triplet excitation of the negatively charged
nitrogen-vacancy center in diamond. In the method, the defect region
is embedded into a pristine environment with an appropriately chosen
cutoff radius. The dynamical matrix of the embedded system is constructed
by sampling the dynamical matrices of the defect-containing supercell
and the pristine-host supercell with an appropriate interaction cutoff radius. 
This construction results in a sparse dynamical matrix, the accuracy of which 
heavily relies on the strongly covalent and nonpolar character of diamond 
leading to rapidly decaying interatomic
force constants. The method improves the vibrational sampling while
also moving towards the dilute-defect limit, suppressing spurious defect-defect
interactions. 
We note that, very similar embedding
workflow has been implemented by Bouquiaux \textit{et~al.} \cite{10.1021/acs.chemmater.3c00537, Bouquiaux_2026} recently.

In recent years, machine-learning interatomic potentials have matured
to the point where quantum-mechanical energies and forces can be reproduced
with great accuracy at a fraction of the cost of density-functional
theory (DFT) calculations \cite{mishinMachinelearningInteratomicPotentials2021,chenUniversalGraphDeep2022,wangScientificDiscoveryAge2023,batatiaFoundationModelAtomistic2025}. Building on this development, Turiansky
\textit{et~al.} recently introduced an approach \cite{turianskyMachineLearningPhonon2026}
in which machine-learning interatomic potentials replace the expensive
first-principles evaluation of force constants required for defect
phonon calculations. Starting from a pretrained foundation model,
the potential is fine-tuned using data generated during routine DFT
geometry relaxations and subsequently used to obtain the dynamical
matrix. By extending the calculations to system sizes beyond those
accessible with direct DFT, the method improves vibrational sampling
and mitigates finite-size effects, similarly to the embedding method.

Both approaches address the computational bottleneck of the Huang-Rhys
framework rather than modifying its underlying physics. Accordingly,
the harmonic, Franck-Condon, equal-frequency, and parallel-mode approximations
remain unchanged, and the associated limitations apply equally to
both methods. In contrast, the MD-ACF approach introduced by Linderälv
\textit{et~al.} \cite{linderalvOpticalLineShapes2025b} follows a
fundamentally different strategy, obtaining the optical lineshape
from autocorrelation functions sampled directly along molecular-dynamics
(MD) trajectories rather than from a static normal-mode decomposition.
This approach enables the capture of anharmonic effects on the spectra. A fundamental
limitation of the method is the classical treatment of nuclear dynamics,
which yields classical rather than quantum correlation functions and
limits the method's accuracy at low temperatures, where nuclear quantum
effects become important.

In this work, we pursue a complementary strategy that remains within
the quantum-mechanical Huang-Rhys framework but addresses its numerical
limitations from a different perspective. Building on the force-based
formulation introduced by Alkauskas \textit{et~al.} \cite{alkauskasFirstprinciplesTheoryLuminescence2014},
we exploit the observation that the transition-induced force associated
with an optical excitation is substantially more spatially confined
than the corresponding structural relaxation. We therefore obtain
the electron-phonon coupling from a comparatively small first-principles
supercell and use phonon interpolation to sample the resulting vibrational
response on dense reciprocal-space grids. This separates the first-principles
treatment of the local defect response from the large-system vibrational
sampling required for converged optical spectra.

Here, we develop this interpolation strategy for the calculation of
Huang-Rhys spectral densities and optical lineshapes of point defects.
The approach provides access to vibrational
sampling far beyond that of the underlying electronic-structure calculation
without requiring the explicit construction and diagonalization of
correspondingly large defect dynamical matrices. We demonstrate the
method for the negatively charged nitrogen-vacancy center in diamond 
and assess its convergence,
accuracy, and computational efficiency. Importantly, the interpolation
step is complementary to embedding and machine-learning-based approaches
and, in principle, can be combined with either to further improve
the sampling of the vibrational continuum. The resulting framework
therefore provides a general and readily applicable route to highly
resolved and converged first-principles optical spectra while retaining
the quantum-mechanical description of lattice vibrations inherent
to Huang-Rhys theory.

\section*{Results}

\subsection*{Optical lineshape theory}

In general, optical transitions between two states can be characterized
by the energy-dependent transition-rate functions $W_{\text{em}}\left(\hbar\omega\right)$
and $W_{\text{abs}}\left(\hbar\omega\right)$ for emission and absorption,
respectively, where the energy in the function arguments is expressed
through the reduced Planck constant $\hbar=h/\left(2\pi\right)$ and
the angular frequency $\omega$. For absorption, however, it is conventional
to use the photon-flux-normalized absorption cross section $\sigma_{\text{abs}}\left(\hbar\omega\right)$
rather than $W_{\text{abs}}\left(\hbar\omega\right)$ itself. Experimentally
determining the absolute magnitude of these quantities is generally
challenging because the measured signal depends on setup- and sample-specific
factors, including photon-collection and detector efficiencies, optical
losses, scattering, and reabsorption. Consequently, optical spectra
are commonly reported in terms of the normalized lineshape functions
$L_{\text{em/abs}}\left(\hbar\omega\right)$, for which the unknown
overall intensity scale is eliminated.

Under the electric-dipole approximation, the light-matter interaction
Hamiltonian reduces to $\hat{H}_{\text{int}}=-\hat{\mathbf{d}}\cdot\hat{\mathbf{E}}$,
the product of the electric dipole moment and electric field operators.
Furthermore, using the Born-Oppenheimer and Condon approximations
\cite{bornZurQuantentheorieMolekeln1927,condonTheoryIntensityDistribution1926},
the lineshape functions take the form 
\begin{equation}
L_{\text{em}}\left(\hbar\omega\right)\propto W_{\text{em}}\left(\hbar\omega\right)\propto\omega^{3}A_{e\rightarrow g}\left(\hbar\omega\right),\label{eq:lineshape-em}
\end{equation}
\begin{equation}
L_{\text{abs}}\left(\hbar\omega\right)\propto\sigma_{\text{abs}}\left(\hbar\omega\right)\propto\omega A_{g\rightarrow e}\left(\hbar\omega\right),\label{eq:lineshape-abs}
\end{equation}
 where $A_{i\rightarrow f}\left(\hbar\omega\right)$ is the spectral
function describing the transition between the thermally populated vibrational manifold of the
initial state $i$, and the vibrational manifold of the final state $f$. This spectral
function can be expressed using the Fermi golden rule as 
\begin{equation}
A^{(\pm)}_{i\rightarrow f}\left(\hbar\omega\right)=\sum_{n,m}p_{i,n}\left|\left\langle \chi_{i,n}\vert\chi_{f,m}\right\rangle \right|^{2}\delta\left(\hbar\omega\pm\left(E_{i,n}-E_{f,m}\right)\right),\label{eq:fermi-golden-rule}
\end{equation}
 where $n,m$ label the $\left|\chi_{i,n}\right\rangle $ and $\left|\chi_{f,m}\right\rangle $
vibrational states on the initial- and final-state potential-energy
surfaces having $E_{i,n}$ and $E_{f,m}$ eigenenergies, while
\begin{equation}
p_{i,n}=\frac{e^{-\beta\left(E_{i,n}-E_{i,0}\right)}}{\sum_{k}e^{-\beta\left(E_{i,k}-E_{i,0}\right)}}=\frac{e^{-\beta E_{i,n}}}{\sum_{k}e^{-\beta E_{i,k}}}
\end{equation}
 is the thermal population of the initial state, where $\beta=\left(k_{\textnormal{B}}T\right)^{-1}$
is the usual thermodynamic beta defined through the Boltzmann constant
$k_{\textnormal{B}}$ and the thermodynamic temperature $T$. The
$\pm$ sign in the $\delta$ distribution of Eqn.~(\ref{eq:fermi-golden-rule})
is negative for emission and positive for absorption, resulting in
photon energies satisfying $\hbar\omega=E_{i,n}-E_{f,m}$ and $\hbar\omega=E_{f,m}-E_{i,n}$,
respectively. 

The difficulty in the \textit{ab initio} evaluation of the spectral
function lies in the overlap of vibrational states $\left\langle \chi_{i,n}\vert\chi_{f,m}\right\rangle $.
This problem is simplified via approximations to arrive at a closed
form in the Huang-Rhys framework of electron-phonon coupling and the
generating function formalism.

\subsection*{Huang-Rhys framework}

Under the Born-Oppenheimer approximation, which was already used to
arrive at equations (\ref{eq:lineshape-em}) and (\ref{eq:lineshape-abs})
by factorizing the electronic transition dipole moment, the total
wave function factorizes into purely electronic and nuclear contributions.
Consequently, the state-specific nuclear Hamiltonian can be expanded
around the equilibrium nuclear configuration $\mathbf{R}_{s,0}$.
Introducing the mass-weighted displacements 
$\mathbf{x}_{s}=\mathbf{M}^{1/2}\left(\mathbf{R}-\mathbf{R}_{s,0}\right)$
and truncating the expansion of the series
at second order, the harmonic approximation of the nuclear Hamiltonian
becomes 
\begin{equation}
\hat{H}_{\text{nuc},s}=E_{\text{min},s}-\frac{\hbar^{2}}{2}\nabla^{T}_{\mathbf{x_{s}}}\nabla^{\,}_{\mathbf{x_{s}}}+\frac{1}{2}\mathbf{x}^{T}_{s}\mathbf{D}_{s}\mathbf{x}_{s},\label{eq:harmonic-nuclear-hamiltonian}
\end{equation}
 where $E_{\text{min},s}$ is the minimum of the potential-energy
surface in electronic state $s$, and $\mathbf{D}_{s}=\mathbf{M}^{-1/2}\mathbf{\Phi}_{s}\mathbf{M}^{-1/2}$
is the dynamical matrix defined through the diagonal nuclear-mass
matrix $\mathbf{M}$ and the interatomic force-constant matrix 
\begin{equation}
\left[\mathbf{\Phi}_{s}\right]_{a,\alpha;b,\beta}=\left.\frac{\partial^{2}E_{s}\left(\mathbf{R}\right)}{\partial R_{a,\alpha}\partial R_{b,\beta}}\right|_{\mathbf{R}=\mathbf{R}_{s,0}},
\end{equation}
 where $a,b$ and $\alpha,\beta$ label atoms and Cartesian coordinate
components, respectively.

The Hamiltonian in Eq.~(\ref{eq:harmonic-nuclear-hamiltonian}) can
be diagonalized, and takes the form 
\begin{equation}
\hat{H}_{\text{nuc},s}=E_{\text{min},s}+\sum_{\nu}\frac{1}{2}\left[\hat{P}^{2}_{s,\nu}+\Omega^{2}_{s,\nu}\hat{Q}^{2}_{s,\nu}\right],
\end{equation}
 describing independent quantum harmonic oscillators, where the normal
coordinate in mode $\nu$ is $Q_{s,\nu}=\mathbf{e}^{\dagger}_{s,\nu}\mathbf{x}_{s}$,
and the eigenvectors $\mathbf{e}_{s,\nu}$ satisfy the 
\begin{equation}
\mathbf{D}_{s}\mathbf{e}_{s,\nu}=\Omega^{2}_{s,\nu}\mathbf{e}_{s,\nu}\label{eq:phonon-eigenproblem}
\end{equation}
 eigenvalue problem. 

In general, the normal coordinates of the initial and final states
are related through a Duschinsky transformation \cite{duschinskyInterpretationElectronicSpectra1937},
$\mathbf{Q}_{f}=\underline{\underline{\mathbf{J}}}\mathbf{Q}_{i}+\mathbf{K}$,
where $\underline{\underline{\mathbf{J}}}$ mixes normal modes and $\mathbf{K}$
is the displacement between the two potential-energy surface minima.
The full multidimensional Franck-Condon/Doktorov \cite{doktorovDynamicalSymmetryVibronic1977}
treatment required to solve this problem is beyond the scope of the
present work; therefore we adopt the usual parallel-mode, equal-frequency
approximation by setting $J_{\mu,\nu}=\delta_{\mu,\nu}$ and imposing
the $\Omega_{i,\nu}=\Omega_{f,\nu}\equiv\Omega_{\nu}$ constraint.

The overlap of vibrational states is thus simplified to the overlap
of a set of independent, displaced harmonic oscillators
\begin{equation}
\left\langle \chi_{i,n}\vert\chi_{f,m}\right\rangle =\prod_{\nu}\left\langle n_{\nu}\vert m_{\nu}\right\rangle ,
\end{equation}
 where $\left|n_{\nu}\right\rangle $ represents the harmonic oscillator
of mode $\nu$ in eigenstate $n_{\nu}$, with energy $\hbar\Omega_{\nu}\left(n_{\nu}+\tfrac{1}{2}\right)$.
Consequently, the overlap in Eq.~(\ref{eq:fermi-golden-rule}) is
reduced to the product of the overlaps of independent displaced harmonic
oscillator states which depend solely on the dimensionless partial
Huang-Rhys factors $S_{\nu}$, conventionally defined through the
$\Delta Q_{\nu}$ displacement as 
\begin{equation}
S_{\nu}=\frac{\Omega_{\nu}\left|\Delta Q_{\nu}\right|^{2}}{2\hbar}.\label{eq:partial-HR-factor-displacement}
\end{equation}

Although the overlap of vibrational states is considerably simplified
by the approximations discussed above, Eq.~(\ref{eq:fermi-golden-rule})
still requires the evaluation of a double infinite summation. This
can be avoided by replacing the $\delta$ distribution by a formal
representation resulting in a generating function formalism.

\begin{figure*}[t!]
	\centering
	\includegraphics[width=\textwidth]{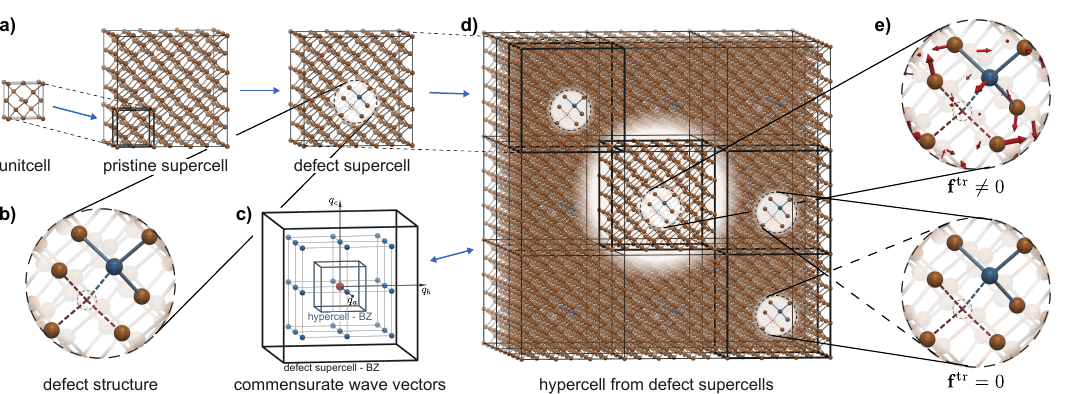}
 	\caption{\label{fig:scaling}\textbf{Construction of the defect supercell and hypercell for phonon interpolation.} \textbf{a)} Regular construction of the defect-containing supercell used in \textit{ab initio} calculations. The defect is embedded into a larger pristine host supercell made from conventional eight-atom cubic diamond unit cells. \textbf{b)} Enlarged defect region within the defect-containing supercell. The NV center is embedded into diamond by replacing a carbon atom (brown sphere) with a nitrogen atom (blue sphere) while also removing an adjacent carbon atom, creating a vacancy (dashed red circle) and dangling bonds (dashed red and blue lines) next to the nitrogen atom. \textbf{c)} Commensurate $\mathbf q$ wave vectors (red ($\mathbf q = 0$) and blue spheres) inside the Brillouin zone of the defect-containing supercell on which the interpolation is performed. This is an equivalent representation of the hypercell. \textbf{d)} The hypercell, constructed by repeating defect-containing supercells. \textbf{e)} Enlarged defect regions within the hypercell. Transition-induced forces are non-zero only in a single reference cell; in all other supercells, the forces are set to zero. This force definition represents an optical transition at a single defect within an array of defects.}
\end{figure*}

\subsection*{Generating function formalism}

By replacing the $\delta$ distribution in Eq.~(\ref{eq:fermi-golden-rule})
with 
\begin{equation}
\delta\left(x\right)=\frac{1}{2\pi\hbar}\int^{+\infty}_{-\infty}e^{-ixt/\hbar}\mathrm{d}t,
\end{equation}
 the spectral function takes the form 
\begin{equation}
A^{(\pm)}_{i\rightarrow f}\left(\hbar\omega\right)=\frac{1}{2\pi\hbar}\int^{+\infty}_{-\infty}e^{-i\left(\omega\pm\Delta\omega_{0}\right)t}G^{\left(\pm\right)}\left(t\right)\mathrm{d}t,
\end{equation}
 where 
\begin{equation}
G^{\left(\pm\right)}\left(t\right)=\prod_{\nu}\sum_{n_{\nu},m_{\nu}}p_{i,n_{\nu}}\left|\left\langle n_{\nu}\vert m_{\nu}\right\rangle \right|^{2}e^{\pm i\left(m_{\nu}-n_{\nu}\right)\Omega_{\nu}t}
\end{equation}
 is the generating function and $\Delta\omega_{0}=\left(E_{i,0}-E_{f,0}\right)/\hbar$.
Expressing the overlap in terms of the $S_{\nu}$ partial Huang-Rhys
factors, the generating function simplifies to 
\begin{equation}
G^{\left(\pm\right)}\left(t\right)=\exp\left\{ \bar{S}\left(\pm t\right)+\bar{S}_{\text{th.}}\left(T;t\right)+\bar{S}_{\text{th.}}\left(T;-t\right)\right\} ,
\end{equation}
 where 
\begin{equation}
\bar{S}\left(t\right)=S\left(t\right)-S\left(0\right)=\sum_{\nu}S_{\nu}\left(e^{i\Omega_{\nu}t}-1\right)
\end{equation}
 is derived from the Fourier transform of the Huang-Rhys spectral
density 
\begin{equation}
S\left(\hbar\omega\right)=\sum_{\nu}S_{\nu}\delta\left(\hbar\omega-\hbar\Omega_{\nu}\right)=\frac{1}{2\pi\hbar}\int^{+\infty}_{-\infty}S\left(t\right)e^{-i\omega t}\mathrm{d}t,\label{eq:huang-rhys-spectral-density}
\end{equation}
 with 
\begin{equation}
S\left(t\right)=\sum_{\nu}S_{\nu}e^{i\Omega_{\nu}t};
\end{equation}
 and 
\begin{equation}
\bar{S}_{\text{th.}}\left(T;t\right)=\sum_{\nu}\bar{n}_{\nu}S_{\nu}\left(e^{i\Omega_{\nu}t}-1\right)
\end{equation}
 is the thermal occupation number weighted version of $\bar{S}\left(t\right)$
with $\bar{n}_{\nu}=\left(e^{\beta\hbar\Omega_{\nu}}-1\right)^{-1}$.
For a more detailed description of theoretical background, we refer
the reader to Section~1 of the Supplementary Information.

\begin{figure*}[t!]
	\centering
	\includegraphics[width=\textwidth]{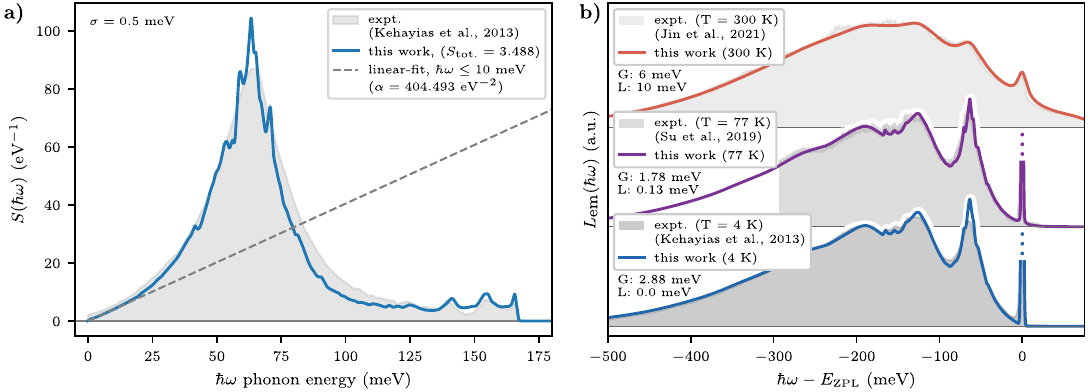}
	\caption{\label{fig:nv-spectras}\textbf{Huang-Rhys spectral density and optical lineshapes of the NV$^-$ center.} \textbf{a)} Calculated Huang-Rhys spectral density for the $^3E\,\rightarrow\,^3A_2$ transition, compared with an experimentally derived reference \cite{kehayiasInfraredAbsorptionBand2013}. Phonon interpolation was performed on a $32\times32\times32$ grid, and a uniform $\sigma=0.5$ meV Gaussian broadening was applied to the $\delta$-distributions in Eq. (\ref{eq:interpolated-HR-spectral-density}). \textbf{b)} Calculated photoluminescence lineshapes at different temperatures, compared with experimental spectra \cite{jinPhotoluminescenceSpectraPoint2021,kehayiasInfraredAbsorptionBand2013,suLuminescenceLandscapesNitrogenvacancy2019}. Additional Gaussian and Lorentzian broadening, with full widths at half maximum (FWHM) \(G\) and \(L\), respectively, was introduced to phenomenologically account for inhomogeneous and homogeneous broadening in the experimental spectra. Temperature-dependent multiphonon contributions are already included through the generating-function formalism.}
\end{figure*}

\subsection*{Interpolation methodology}

Within the approximations adopted above, the vibrational contribution to the optical lineshape is completely specified by the Huang-Rhys spectral density $S\left(\hbar\omega\right)$. As defined in Eq.~(\ref{eq:huang-rhys-spectral-density}),
this quantity is constructed from the phonon frequencies $\Omega_{\nu}$ and the corresponding partial Huang-Rhys factors $S_{\nu}$, with the latter determined by the projection of the mass-weighted equilibrium displacement $\Delta\mathbf{x}$ onto the phonon eigenmodes.
However, in practice, $\Delta\mathbf{x}$ exhibits characteristic long-range behavior~\cite{clouetElasticModelingPointdefects2018} $|\Delta\mathbf{x}|\sim 1/R^{2}$ that is difficult to contain in finite-sized supercells computationally accessible by \textit{ab initio} methods.
Even when the perturbation associated with the defect transition is strongly localized, the resulting structural relaxation extends considerably further into the surrounding lattice.
This extended displacement field associated with a localized defect transition originates from the long-wavelength elastic response of the crystal rather than an equally extended force perturbation. 
Within the harmonic approximation, the relaxation can be expressed in terms of its mass-normalized transition-force source,
\begin{equation}
\mathbf{D}\Delta\mathbf{x}=\mathbf{f}^{\textnormal{tr}}.\label{eq:displacement-force-equivalence}
\end{equation}
The local component of this force source is exponentially localized, with $|\mathbf{f}^{\mathrm{tr}}|\sim e^{-R/\xi}/R$.
This force represents the change in interatomic forces induced by the electronic transition at fixed atomic positions. 
Remarkably, while $\Delta\mathbf{x}$ is governed and modulated by the phonon eigenproblem of the host, $\mathbf{f}^{\mathrm{tr}}$ reflects the change in the electronic potential at fixed atomic positions, and thus carries the raw intrinsic consequence of a localized electronic transition of the defect.
For transitions between localized defect states in a gapped host, the dominant short-range component of the transition-force perturbation is concentrated near the defect, consistent with the nearsightedness of the electronic response \cite{prodanNearsightednessElectronicMatter2005}.
An equivalent force-based representation was previously employed by Alkauskas \textit{et~al.} \cite{alkauskasFirstprinciplesTheoryLuminescence2014}.

In a periodically repeated \textit{ab initio} supercell, the equilibrium displacements consequently contain the elastic response to an entire periodic array of defects. The displacement fields of neighboring images may therefore overlap appreciably even when the local electronic structure is already well converged. In contrast, the more spatially confined transition force can provide a substantially cleaner approximation to the isolated-defect perturbation, allowing the corresponding electron-phonon coupling to be reconstructed using a larger vibrational representation.

Projecting the forces, defined through Eq.~(\ref{eq:displacement-force-equivalence}),
onto the phonon eigenvectors which solve Eq.~(\ref{eq:phonon-eigenproblem})
gives $\Delta Q_{\nu}=\mathbf{e}^{\dagger}_{\nu}\mathbf{f}^{\textnormal{tr}}/\Omega^{2}_{\nu}$.
Substitution into the displacement-based formulation of Eq.~(\ref{eq:partial-HR-factor-displacement})
then gives the equivalent force representation 
\begin{equation}
S_{\nu}=\frac{\left|\mathbf{e}^{\dagger}_{\nu}\mathbf{f}^{\textnormal{tr}}\right|^{2}}{2\hbar\Omega^{3}_{\nu}}.\label{eq:partial-HR-factor-force}
\end{equation}
 Thus, no additional approximation is introduced by replacing the
displacement with the force within the harmonic treatment; the advantage
is instead that the quantity transferred from the finite \textit{ab
initio} calculation is considerably more local.

This locality makes it possible to separate the first-principles description
of the defect from the sampling of its vibrational environment. We
construct a hypercell by periodically repeating the defect-containing
\textit{ab initio} supercell $M_{1}\times M_{2}\times M_{3}$ times,
corresponding to $N_{q}=M_{1}M_{2}M_{3}$ commensurate wave vectors.
Because the force-constant kernel of this periodically repeated system
depends only on the relative separation of supercells, the real-space
hypercell dynamical matrix has a block-circulant structure and is
block-diagonalized by a discrete Fourier transformation, 
\begin{equation}
\mathbf{D}_{a,\alpha;b,\beta}(\mathbf{q})=\sum_{L}\mathbf{D}_{a,\alpha;b,\beta}(0,L)e^{i\mathbf{q}\left[\mathbf{R}_{b}(L)-\mathbf{R}_{a}(0)\right]}.
\end{equation}
 The vibrational problem can therefore be solved independently for
each $\mathbf{q}$, 
\begin{equation}
\mathbf{D}(\mathbf{q})\mathbf{e}_{\nu}(\mathbf{q})=\Omega^{2}_{\nu}(\mathbf{q})\mathbf{e}_{\nu}(\mathbf{q}),
\end{equation}
 rather than by explicitly diagonalizing the full hypercell dynamical
matrix. Here, $\mathbf{e}_{\nu}(\mathbf{q})$ describes the mode within
one defect supercell. The corresponding normalized eigenmode of the
complete hypercell is 
\begin{equation}
E_{\nu;L,a,\alpha}(\mathbf{q})=\frac{1}{\sqrt{N_{q}}}e^{i\mathbf{q}\mathbf{R}_{a}(L)}e_{\nu;a,\alpha}(\mathbf{q}),\label{eq:hypercell-eigenmodes}
\end{equation}
 where $L$ labels the repeated supercells.

Although the vibrational basis is constructed from a periodically
repeated supercell, the optical transition should represent a single
defect rather than a coherent transition of every defect image. We
therefore define the mass-normalized transition force to be non-zero
only in one reference supercell as 
\begin{equation}
\mathbf{f}^{\textnormal{tr}}_{L}=\mathbf{f}^{\textnormal{tr}}\delta_{L,0},\label{eq:hypercell-force-source}
\end{equation}
 where $\delta_{L,0}$ is the Kronecker delta. Applying the force-based
representation of partial Huang-Rhys factors in Eq.~(\ref{eq:partial-HR-factor-force})
to the hypercell modes of Eq.~(\ref{eq:hypercell-eigenmodes}) and
substituting Eq.~(\ref{eq:hypercell-force-source}) gives 
\begin{equation}
S_{\mathbf{q},\nu}=\frac{\left|\sum_{L,a}\mathbf{E}^{\dagger}_{\nu;L,a}(\mathbf{q})\:\mathbf{f}^{\textnormal{tr}}_{L;a}\right|^{2}}{2\hbar\Omega^{3}_{\nu}(\mathbf{q})}=\frac{\left|\sum_{a}\mathbf{e}^{\dagger}_{\nu;a}(\mathbf{q})\:\mathbf{f}^{\textnormal{tr}}_{a}e^{-i\mathbf{q}\mathbf{R}_{a}(0)}\right|^{2}}{2N_{q}\hbar\Omega^{3}_{\nu}(\mathbf{q})}\text{,}
\label{eq:hypercell-partial-HR-factors}
\end{equation}
 where the $1/N_{q}$ factor follows directly from the normalization
of the hypercell eigenmodes. Increasing the hypercell size therefore
leaves the reference transition-force source unchanged while providing
an increasingly dense set of phonon frequencies and eigenvectors.
In this context, the partial Huang-Rhys factors are therefore labeled by a
$\left(\mathbf{q},\nu\right)$ pair instead of a single hypercell mode
index. Consequently, the construction of the Huang-Rhys spectral density
involves a summation over the commensurate wave vectors, 
\begin{equation}
S\left(\hbar\omega\right)=\sum_{\mathbf{q},\nu}S_{\mathbf{q},\nu}\,\delta\left(\hbar\omega-\hbar\Omega_{\nu}(\mathbf{q})\right).\label{eq:interpolated-HR-spectral-density}
\end{equation}
 In this way, the local electron-phonon coupling remains determined
by the tractable ab initio supercell, whereas the extended elastic
response and the detailed vibrational continuum are recovered through
phonon interpolation.

\section*{Discussion}

\subsection*{The nitrogen-vacancy center in diamond}

We demonstrate and discuss the capabilities of our interpolation methodology on the negatively charged nitrogen-vacancy (NV$^{-}$) center in diamond.
NV$^{-}$ is a paradigmatic solid-state point defect with numerous applications
in quantum technologies \cite{pompiliRealizationMultinodeQuantum2021,dohertyNitrogenvacancyColourCentre2013,jelezkoObservationCoherentOscillations2004,gruberScanningConfocalOptical1997,laddQuantumComputers2010,awschalomQuantumSpintronicsEngineering2013,galiInitioTheoryNitrogenvacancy2018}. Its primary optical transition ($^{3}E\,\rightarrow\,^{3}A_{2}$)
has a zero-phonon line at approximately 1.945~eV (637 nm) and exhibits
strong electron-phonon coupling, resulting in a prominent phonon
sideband. The sideband is dominated by a broad contribution associated
with acoustic phonons, accompanied by weaker features near the optical-phonon
band edge. Resolving these contributions requires accurate sampling
of electron-phonon coupling across a wide range of phonon energies,
making the NV$^{-}$ center a suitable benchmark for the interpolation
method introduced here.

Fig.~\ref{fig:nv-spectras} shows the Huang-Rhys spectral density
and photoluminescence lineshapes of the NV$^{-}$ center calculated
using the interpolation method introduced here, together with
experimental references. 

The calculated spectral density reproduces the dominant broad contribution
centered around 63~meV, as well as the finer features near the optical-phonon
band edge above 130~meV. Furthermore, the linear behavior of the
spectral density in the low-energy acoustic region is recovered. These
results demonstrate that the interpolation method resolves electron-phonon
coupling contributions across the phonon spectrum, including both
acoustic and optical modes. At a $4\times4\times4$ supercell size (corresponding
to 512 carbon atoms for pristine diamond) and $32\times32\times32$ hypercell size, we calculate a total Huang-Rhys
factor of $S_{\textnormal{tot.}}=3.49$, and a linear slope of $\alpha\approx404.5$~eV$^{-2}$
in the low-energy acoustic regime.

Using the generating-function formalism, we subsequently calculate
the temperature-dependent photoluminescence lineshapes, which reproduce
the main features of the experimental spectra after accounting for
additional broadening. Nevertheless, exact agreement is not expected
within the approximations employed. In particular, at elevated
temperatures, deviations may arise from anharmonic effects that are neglected
within the harmonic approximation. More generally, the parallel-mode and equal-frequency approximations may also result in further deviations between theory and experiment.

\subsection*{Locality of short-range transition-force sources}

The accuracy of the interpolation method relies on the spatial localization of the transition-induced force source. 
In particular, the force distribution must be sufficiently localized within the first-principles reference supercell to ensure convergence of the calculated electron-phonon coupling.
Indeed, transition-induced force for NV$^-$ is exponentially localized by an $|\mathbf{f}^{\mathrm{tr}}(r)|\sim e^{-r/\xi}/r$ envelope,
where the characteristic localization length is $\xi\approx 3$~\AA~(see Section 4 in Supplementary Information). 
Numerically, the $7.1$~\AA~lattice length of our smallest $2\times2\times2$ supercell is too small to accommodate a sphere of forces exhibiting a diameter $2\xi = 6$~\AA~convergently.
Illustratively, walking just $3.55$~\AA~from the defect's center one would approach the supercell edge already, where the characteristic exponential quenching nature of the forces is only about to start.

On the other hand, transition forces start to overlap on the periodic-image boundaries that may suppress or create spurious effects on the localized force distribution near the boundaries.
Such finite-size effects become particularly important when the reference-supercell dimensions are comparable to the spatial extent of the localized transition-force source, leaving a significant fraction of the electron-phonon interaction to be truncated in our smallest model (at least $\bar{C}(3.55$~\AA$)$~$\sim 10$\% measured in squared forces), see Fig.~S2(b) in the Supplementary for details.
Indeed, the total Huang-Rhys factor for our smallest supercell is an overly quenched $S_\mathrm{tot}=\sum_{\mathbf{q},\nu} S_{\mathbf{q},\nu} = 2.42$ value.
In contrast, larger $3\times3\times3$ and $4\times4\times4$ cases do provide significantly more convergent results: $S_\mathrm{tot}=3.37$ and 3.49 respectively.
On the largest supercell we crop only $\bar{C} (7.1$~\AA$)$~$\sim 0.1$\% of the force squares.
More importantly, this $0.1$\% value can be identified as a tentative relative error estimator for $S_\mathrm{tot}=3.49$ because partial Huang-Rhys factors ($S_{\mathbf{q},\nu}$) are proportional to force squares see Eq.~\eqref{eq:hypercell-partial-HR-factors}.
In reality, the relative error for the $3\times3\times3$ case is approximately $\frac{3.37-3.49}{3.49} \approx 3.4$\% by interpreting the $4\times4\times4$ system as the converged bulk limit, which is significantly larger than the tentative relative error naively estimated from $\bar{C}(5.3$~\AA$)$~$\sim 1$\%.
Therefore, an exponential decay trend ($\approx\frac{1}{7}$ factor upon enlarging the supercell by one step) exists that limits the relative error for $S_\mathrm{tot}$ in large supercells as: $\delta S_{tot}/S_{tot} \propto \bar{C}(r) \propto e^{r/\lambda}$ where $\lambda\approx  0.906$~\AA. 
Taking this into account, we can estimate the relative error this way for the  $2\times2\times2$ case which is $7\times3.4$\%~$\approx$~$24$\% that tentatively estimates its $S_\mathrm{tot} \approx3.49\times(1-0.28)=2.64$ correctly close to its \textit{ab initio} 2.42 value.
Therefore, we conclude that \textit{ab initio} simulation in excessively large supercells is not required for transition forces because they are exponentially confined around the defect.
Assuming the observed approximately eightfold reduction in the finite-size error continues, we estimate a residual relative error of approximately 0.4\% for the $4\times4\times4$ supercell.

Another important point is to clarify the origin of spurious long range effects in the displacement field.
The Fourier transform of the transition force is bound by the ($|\mathbf{f}^{\mathrm{tr}}(q)|\sim1/(q^{2}+\xi^{-2})$) envelope which will admit a regular analytical expansion around $\mathbf{q}=0$ for $\mathbf{f}^{\mathrm{tr}}(\mathbf{q})$.
Translational invariance constrains the net force to vanish, eliminating the zeroth-order ($\xi$) constant term.
Therefore, the leading contribution will be proportional to $\mathbf{q}$, where the proportionality constant is the $P_{i,j}^{\mathrm{tr}}=\sum_{a}R_{a,j}F_{a,i}^{\mathrm{tr}}$ elastic-dipole moment \cite{varvenneElasticDipolesPoint2017} (elastic dipole tensor) of the transition force distribution, where  $\mathbf{F}_{a}^{\mathrm{tr}}$ is the transition force acting on the $a$-th nucleus at position $\mathbf{R}_{a}$.
Thus, the transition forces can be expanded as $\mathbf{f}^{\mathrm{tr}}(\mathbf{q})\sim\mathbf{P}^{\mathrm{tr}}\mathbf{q}$, thus in short: the transition forces scale linearly in $q$: $|\mathbf{f}^{\mathrm{tr}}(\mathbf{q})|\propto q$ at the long-wavelength acoustic limit, see the Supplementary Information for details.
We note that the $P_{i,j}$ is defined as the forces exerted on the host atoms due to the initial (unrelaxed) placement of a defect, or more precisely, defined through elastic interaction energy between an external strain field and the point defect~\cite{varvenneElasticDipolesPoint2017, varvennePointDefectModeling2013, Bacon_1980, PhysRevB.87.144107, PhysRevB.94.054106, Vattr_2016}.
Here, we are extending this notation for transition forces, the forces appearing upon exciting a defect to one of its electronically excited states.

Meanwhile, the acoustic eigenvalues of the dynamical matrix scale as $\Omega^2(\mathbf{q})\propto q^2$ as $\mathbf{q}\rightarrow0$ as the dispersion relation of acoustic phonons is linear: $\Omega_\lambda(\mathbf{q})=c_\lambda({\hat{\mathbf{q}}})\times q$ where the proportionality parameter is the speed of sound ($c_\lambda(\hat{\mathbf{q}})$) of the host for $\lambda$ polarization index (one longitudinal, and two transverse) and  ${\hat{\mathbf{q}}}={\mathbf{q}}/{q}$ direction.
Consequently, following from Eq.~\eqref{eq:displacement-force-equivalence}, the corresponding displacement response will behave as $\Delta x(\mathbf{q})\propto q^{-1}$, giving rise to the characteristic $r^{-2}$ decay of the elastic-dipole displacement field in three dimensions \cite{clouetElasticModelingPointdefects2018}.
The phonon density of states for three dimensional materials is $g(\omega)\propto\omega^{2}$ in the acoustic regime (where $\omega \propto q$ and the proportionality factor is the ($c$) speed of sound),
Eq.~\eqref{eq:partial-HR-factor-displacement} leads to:
\begin{equation}
S(\omega)\propto\underset{\propto q^{2}}{\underbrace{g(\omega)}}\times\underset{\propto q}{\underbrace{\Omega({\mathbf{q}})}}\times\underset{\propto q^{-2}}{\underbrace{(\Delta x(\mathbf{q}))^{2}}}\propto\omega \text{.}
\end{equation}
Thus, we can postulate that for a spatially localized transition-force distribution with a finite, nonzero first spatial moment, the acoustic force amplitude is linear in the phonon wave vector in the long-wavelength limit. 
Combined with the quadratic acoustic dynamical-matrix eigenvalues and the three-dimensional phonon density of states, this yields a Huang-Rhys spectral density that increases linearly with phonon energy, $S(\hbar\omega)=\alpha\times\hbar\omega$ as $(\hbar\omega\rightarrow0)$, as observed in Fig.~\ref{fig:nv-spectras}.
If the leading force moment vanishes by symmetry or cancellation, higher-order terms determine the low-energy behavior.

\subsection*{Insignificance of long-range transition-force}

However, one may approach the matter from a completely different point of view.
Alternatively, the result of the electronic transition can be viewed from the perspective of the displaced electronic charge.
In that context, the redistributed charge density additionally generates a long-range electric field that can also couple to the host ions.
The resulting bare electrostatic field decays algebraically as $1/r^{L+2}$, where $L$ denotes the lowest nonvanishing multipole order of the $\Delta n(\mathbf{r})$ transition-induced charge-density difference.
In our charge-conserving optical excitation the zeroth-order monopole term vanishes. 
Thus, the leading term for NV will be the change in its dipole moment: $|\Delta\mathbf{d}|=4.33$~Debye as reported in Ref.~\cite{galiInitioTheoryNitrogenvacancy2018}.
This dipole moment would induce a long-range transition-induced electric field that scales as $1/r^3$.
However, this bare contribution cannot couple to acoustic phonon modes directly.
Born effective charges within a unit cell satisfy an acoustic sum rule, which removes the rigid-translation response to a uniform electric field, but does not eliminate spatial-gradient and flexoelectric acoustic coupling~\cite{tagantsevPiezoelectricityFlexoelectricityCrystalline1986, hongFirstprinciplesTheoryCalculation2013, stengelFlexoelectricityDensityfunctionalPerturbation2013}.
Thus, the leading term of that response will scale as $1/r^4$ for hosts without inversion symmetry through piezoelectric coupling.
However, diamond is a crystal exhibiting inversion symmetry thus the long-range force will decay more rapidly, as $1/r^5$ within the assumed flexoelectric model.
Thus, in principle that $1/r^5$ could overtake the $e^{-r/\xi}/r$ local transition forces.
Within the illustrative flexoelectric model developed in the Supplementary Information, the extrapolated algebraic contribution overtakes the assumed exponentially localized force envelope at approximately $\sim$40~\AA.
However, the residual cumulative force weight is $1-C($35-42~\AA$)\sim10^{-12}$ at the crossover point, thus one can conclude that the transition forces induced by the long range fields are negligible.
Thus, we conclude that the applicability of the interpolation method is governed by the convergence of the highly localized transition-force source within the reference supercell.

\subsection*{Remarks on the computational footprint}

For a fixed reference supercell, the computational cost of pysideband scales linearly with the number of phonon $\mathbf{q}$ points, whose independent treatment enables straightforward parallelization.
By accumulating the spectral density without retaining eigenvectors from previously processed wave vectors, the working-memory requirement per worker remains essentially independent of the interpolation-grid size.
In our NV-center calculations, interpolation is substantially less demanding than the HSE geometry relaxations or the PBE finite-displacement force-constant calculations for the 512-atom reference supercell.
Only for the densest $32\times32\times32$ grid does the computational cost approach, while still remaining below, that of the first-principles calculations.
We note that our code relies on the same \textit{ab initio} inputs that other available codes: \textit{PyPhotonics}~\cite{Tawfik_2022}, 
\textit{PyPL}~\cite{jinPhotoluminescenceSpectraPoint2021, Jin_2022, 10.1063/5.0349651}, \textit{DefectPL}~\cite{PhysRevB.111.104109, lsxj-nvhw}, 
\textit{lineshape\_tools}~\cite{turianskyMachineLearningPhonon2026},
\textit{Lumabi}~\cite{Bouquiaux_2026}
provide complementary first-principles lineshape workflows.
Our present method introduces a new reciprocal-space interpolation strategy for evaluating the coupling of localized transition forces to densely sampled phonon modes.
By interpolating the vibrational response of a DFT-accessible 512-site reference supercell, the method resolves detailed phonon-sideband features corresponding to an effective hypercell containing more than 16 million atoms, without requiring additional large-supercell DFT calculations.
Unlike conventional embedding approaches\cite{ alkauskasFirstprinciplesTheoryLuminescence2014, Bouquiaux_2026}, which combine defect-region and pristine-host interatomic force constants, our method directly uses the reference-supercell force constants without constructing an explicitly embedded defect-bulk composite dynamical matrix. 
Although recent implementations have substantially automated the embedding \cite{Bouquiaux_2026}  workflow, structural mapping and the combination of defect and bulk force constants remain additional procedural steps. 
Our approach avoids these steps, simplifying the computational workflow while retaining dense vibrational sampling and offer linear computational scaling $\mathcal{O}(Nn^{2})$ in target system size ($N$) and reference system size ($n$) as opposed to embedding methods that are limited by matrix diagonalization of their embedded defect-bulk composite dynamical matrix that scales as $\mathcal{O}(N^{3})$.
For further details on the computational scaling, see Fig.~\ref{figure1} and the Supplementary Information.

\subsection*{Summary}

In this work, we have developed a phonon-interpolation method for
calculating Huang-Rhys spectral densities and optical lineshapes of
point defects from first-principles calculations. By exploiting the
spatial localization of transition-induced forces, the method separates
the description of local electron-phonon coupling from the extended
vibrational response of the host, enabling dense phonon sampling without
explicitly constructing and diagonalizing large hypercell dynamical
matrices. Applied to the negatively charged nitrogen-vacancy center
in diamond, the method resolves the dominant acoustic contribution,
the finer features in the optical-phonon region, and the low-energy
behavior of the Huang-Rhys spectral density. The resulting temperature-dependent
photoluminescence lineshapes reproduce the main features of experimental
spectra after accounting for additional broadening. The accuracy of
the method relies on the convergence of the transition-force source
within the first-principles supercell, which may require particular
care in materials exhibiting long-range electrostatic interactions.
Furthermore, the interpolation approach is complementary to existing
embedding and machine-learning methods and can, in principle, be combined
with either to further improve vibrational sampling.

\section*{Methods}

\subsection*{DFT calculations}

Ab initio calculations were performed using the Vienna \textit{Ab
initio} Simulation Package (VASP) \cite{kresseEfficientIterativeSchemes1996,kresseInitioMolecularDynamics1993a}.
In all calculations spin-polarized DFT was used together with projector-augmented-wave
(PAW) potentials \cite{kresseUltrasoftPseudopotentialsProjector1999,blochlProjectorAugmentedwaveMethod1994}
to reduce the necessary plane-wave basis size. State-specific
equilibrium geometries were obtained using the screened hybrid Heyd-Scuseria-Ernzerhof
(HSE) functional \cite{heydHybridFunctionalsBased2003,krukauInfluenceExchangeScreening2006},
in which 25\% of screened Fock exchange is admixed to the semilocal
exchange of the Perdew-Burke-Ernzerhof (PBE) \cite{perdewGeneralizedGradientApproximation1996}
functional. Vibrational calculations were performed using the computationally
less expensive PBE functional, which has been shown to provide vibrational
properties of comparable accuracy for similar systems \cite{razinkovasVibrationalVibronicStructure2021,hummerHeydScuseriaErnzerhofHybridFunctional2009}.
Vibrational phonon-modes were obtained using the finite-differences
method implemented in VASP, and \textit{phonopy} \cite{togoFirstprinciplesPhononCalculations2023,togoImplementationStrategiesPhonopy2023}
was used to extract force-constants.

For the NV$^{-}$ center, a 500 eV cutoff energy was used and geometries
were optimized until the force on each ion fell below $5\cdot10^{-3}$
eV/\AA. The $^{3}A_{2}$ ground-state structure was obtained by restricting
the total spin-channel occupations to differ by 2, resulting in the
$a^{2}_{1}e^{\uparrow}_{x}e^{\uparrow}_{y}$ defect Kohn-Sham (KS)
orbital occupation. The $^{3}E$ excited-state geometry was obtained
using the \textgreek{Δ}SCF method, with fixed orbital occupations.
Specifically, the highest-energy Kohn-Sham electron in the spin-minority
channel was promoted to the next available orbital, yielding the $a^{\uparrow}_{1}e^{2}_{x}e^{\uparrow}_{y}$
defect KS-orbital occupation.

\subsection*{Calculation of spectral functions}

All calculated spectral functions and optical lineshapes were obtained
using the \textit{pysideband} command-line Python software developed
by the authors, which implements the interpolation method introduced
in the present work. The complete computational workflow is summarized
by the following sequence: \begin{equation}
	\begin{tikzcd}[column sep=normal, row sep=normal]
		\Delta\mathbf R
		    \arrow[r]
		&
		\mathbf F^{\textnormal{tr}}
		    \arrow[r]
		&
		S_{\mathbf q\nu}
		    \arrow[d]
		    \arrow{dr}{K_G}
		&
		\mathbf D(\mathbf q)
			\arrow[l]
		\\
		L_{\mathrm{em/abs}}(\hbar\omega)
		&
		A_{\mathrm{em/abs}}(\hbar\omega)
		    \arrow[l]
		&
		G(t)
		    \arrow{l}[swap]{K_G,\,K_L}
		&
		S(\hbar\omega)
		    \arrow[l, dashed]
	\end{tikzcd}
\end{equation}where ab initio data is supplied to form $\Delta\mathbf{R}$ and $\mathbf{D}(\mathbf{q})$,
and empirical smearing parameters are supplied for Gaussian $K_{G}$
and Lorentzian $K_{L}$ kernels, on top of specified target temperatures,
which already broaden spectral features. The implementation uses functionalities
provided by the open-source Python package \textit{phonopy} \cite{togoFirstprinciplesPhononCalculations2023,togoImplementationStrategiesPhonopy2023},
including the handling of force-constant data from different first-principles
codes and the efficient construction and diagonalization of wave-vector-dependent
dynamical matrices $\mathbf{D}(\mathbf{q})$.

All NV$^{-}$ center related spectra shown here were calculated on
a $32\times32\times32$, $\Gamma$-centered interpolation grid based
on ab initio calculations made on a $4\times4\times4$ conventional
diamond supercell, containing 512 atomic sites (511 atoms after introducing
the NV$^{-}$ defect). In total, 16,744,448 atoms are present in the final system (hypercell),
where vibrational modes are calculated. In Fig.~\ref{fig:nv-spectras}a a $\sigma=0.5$
meV Gaussian smearing was applied to the Huang-Rhys spectral density
$S\left(\hbar\omega\right)$. In Fig.~\ref{fig:nv-spectras}b, optical
photoluminescence lineshapes are shown for 4~K, 77~K and 300~K
temperatures. For each target temperature, thermal effects are included
through the generating-function formalism. Additional Gaussian and
Lorentzian broadening kernels, $K_{G}$ and $K_{L}$, are subsequently
applied to account for broadening mechanisms not explicitly included
in the calculation. Values for the broadening were determined to best
fit experimental data, together with a fitted smooth phenomenological baseline for
the 77~K experimental data to account for the decrease in detector
sensitivity. The experimentally measured zero-phonon-line energy of
1.945~eV was used when evaluating the $\omega^{3}$ photon-energy
weighting of the emission lineshape.

\section*{Data availability}

The input files required to reproduce the spectra presented in this work using the \textit{pysideband} software are available at {[}REPOSITORY/DOI{]}.
These include the state-specific atomic structures in VASP POSCAR format, the force-constant data compatible with \textit{phonopy}, and the corresponding input files for the \textit{pysideband} software. 
VASP is a commercial program and is available from their distributors under license. 

\section*{Code availability}

The implementation of the phonon-interpolation method introduced in
this work is available as a command-line Python software, \textit{pysideband}.
The software can be installed from the official Python Package Index
(PyPI) via '\verb|pip install pysideband|'. The source code is also publicly hosted on a GitHub repository \cite{pysideband-github}.
We emphasize that our code is not limited to the VASP \textit{ab initio} package.
In principle, the workflow can make use of any electronic-structure code, especially those for which \textit{phonopy} offers an automated interface to generate the FORCE\_CONSTANTS file.

\section*{Acknowledgments}

G. T. was supported by the J\'anos Bolyai Research Scholarship of the Hungarian Academy of Sciences and by NKFIH (National Research, Development and Innovation Office of Hungary) under Grant No. STARTING 150113. 
G. T. and Z. S. acknowledge access to high-performance computational resources awarded by Digital Government Development and Project Management Ltd. at the Komondor HPC facility based in Hungary.
\section*{Author contributions}
Z. S. performed the computations and developed and maintains the entire pysideband software, which is based on a conceived idea and an initial proof-of-concept implementation developed by G. T. The analysis of the results and code testing was carried out by both authors. 
Z. S. prepared all figures for the main manuscript and also for the Supplementary Information.
All authors provided critical feedback and helped shape the research, analysis and writing the manuscript where Z. S. was supervised by G. T.

\section*{Competing interests}
The authors declare no competing interests.

\bibliographystyle{unsrt}
\bibliography{phonon_interpolation_npj}

\clearpage
\newpage
\setcounter{page}{1}
\setcounter{equation}{0}
\setcounter{figure}{0}
\let\oldtheequation\theequation
\renewcommand{\theequation}{S\oldtheequation}

\setcounter{table}{0}
\let\oldthetable\thetable
\renewcommand{\figurename}{Supplementary Figure}
\renewcommand{\tablename}{Supplementary Table}

\renewcommand{\thefigure}{\arabic{figure}}
\renewcommand{\thetable}{\arabic{table}}

\renewcommand{\theequation}{S\arabic{equation}}
\end{multicols}
	
\part*{\begin{center}Supplemental Material\end{center}}

\section{Theoretical background}

\subsection{Optical transitions and vibronic lineshape functions}

Consider an optical transition between two vibronic states, $\left|i,n\right\rangle \rightarrow\left|f,m\right\rangle $,
where $n$ and $m$ denote the occupied vibronic levels of the initial
$i$ and final $f$ electronic states, respectively. The transition
probability per unit time can be expressed through Fermi's golden
rule \cite{diracQuantumTheoryEmission1927,zhangFermisGoldenRule2016}
as 
\begin{equation}
\Gamma_{i,n\rightarrow f,m}=\frac{2\pi}{\hbar}\left|\left\langle i,n\left|\hat{H}_{\textnormal{int}}\right|f,m\right\rangle \right|^{2}\rho(E_{f}),\label{eq:Fermi's-golden-rule}
\end{equation}
 where $\hbar$ is the usual reduced Planck constant, $\hat{H}_{\textnormal{int}}$
is the light-matter interaction Hamiltonian driving the optical transition
and $\rho(E_{f})$ is the density-of-states (DOS) of the final-state
vibronic manifold at energy $E_{f}$.
Under the electric-dipole approximation, 
\begin{equation}
\hat{H}_{\textnormal{int}}=-\hat{\mathbf{d}}\cdot\hat{\mathbf{E}},
\end{equation}
 where $\hat{\mathbf{d}}$ and $\hat{\mathbf{E}}$ are the electric-dipole
and electric-field operators, respectively. For an isotropic, non-dispersive
dielectric medium with refractive index $n_{r}$, the quantized electric
field can be written as 
\begin{equation}
\hat{\mathbf{E}}(\mathbf{r},t)=i\sum_{\mathbf{k},\lambda}\sqrt{\frac{\hbar\omega_{\mathbf{k}}}{2\epsilon_{0}n^{2}_{r}V}}\left[\mathbf{e}_{\mathbf{k},\lambda}\hat{a}_{\mathbf{k},\lambda}e^{i\left(\mathbf{k}\cdot\mathbf{r}-\omega_{\mathbf{k}}t\right)}-\mathbf{e}^{*}_{\mathbf{k},\lambda}\hat{a}^{\dagger}_{\mathbf{k},\lambda}e^{-i\left(\mathbf{k}\cdot\mathbf{r}-\omega_{\mathbf{k}}t\right)}\right],\label{eq:quantized-electric-field}
\end{equation}
 with quantization volume $V$, polarization vector $\mathbf{e}_{\mathbf{k},\lambda}$,
and photon creation and annihilation operators $\hat{a}^{\dagger}_{\mathbf{k},\lambda}$
and $\hat{a}_{\mathbf{k},\lambda}$ \cite{huttnerQuantizationElectromagneticField1992}.
Under the electric-dipole approximation, the spatial extent of optical center at $\mathbf{r}_{0}$ is much smaller than the optical wavelength, so
the $e^{i\mathbf{k}\cdot\mathbf{r}}$ phase factor can be treated as a constant
across the optical center, effectively yielding $e^{i\mathbf{k}\cdot\mathbf{r}}\approx1$. 

Within the Born-Oppenheimer approximation \cite{bornZurQuantentheorieMolekeln1927},
the vibronic states can be factorized into an electronic wave function,
which parametrically depends on the nuclear coordinates, collectively
represented by $\mathbf{R}$, and a nuclear wave function evolving
on the corresponding adiabatic potential energy surface (APES), 
\begin{equation}
\left|s,n\right\rangle =\left|s;\mathbf{R}\right\rangle \left|\chi_{s,n}\right\rangle ,\qquad s\in\left\{ i,f\right\} ,
\end{equation}
 where $n$ labels the vibrational state on electronic surface $s$.
The corresponding total vibronic energy is denoted $E_{s,n}$, and,
in particular, the zero-phonon-line (ZPL) is the energy difference
between the lowest vibronic levels of the two electronic manifolds,
\begin{equation}
E_{ZPL}=\left|E_{f,0}-E_{i,0}\right|.
\end{equation}

However, the vibronic states of the optical center are coupled to a
quantized photon field, and the optical transition involving a photon
with wave vector $\mathbf{k}$ and polarization direction $\lambda$
becomes $\left|i,n;N_{\mathbf{k},\lambda}\right\rangle \rightarrow\left|f,m;N_{\mathbf{k},\lambda}\pm1\right\rangle $,
depending on whether a photon is emitted or absorbed during the transition.
Under the present approximations, this coupling may simply be expressed
as 
\begin{equation}
\left|s,n;N_{\mathbf{k},\lambda}\right\rangle =\left|s,n\right\rangle \otimes\left|N_{\mathbf{k},\lambda}\right\rangle =\left|s;\mathbf{R}\right\rangle \left|\chi_{s,n}\right\rangle \otimes\left|N_{\mathbf{k},\lambda}\right\rangle ,
\end{equation}
 treating the vibronic and photonic degrees of freedom as independent, and the $\hat{a}^{\dagger}_{\mathbf{k},\lambda}$
and $\hat{a}_{\mathbf{k},\lambda}$ creation and annihilation operators
act on the photon number states $\left|N_{\mathbf{k},\lambda}\right\rangle $
with the usual bosonic rules, 
\begin{equation}
\hat{a}^{\dagger}_{\mathbf{k},\lambda}\left|N_{\mathbf{k},\lambda}\right\rangle =\sqrt{N_{\mathbf{k},\lambda}+1}\left|N_{\mathbf{k},\lambda}+1\right\rangle ,
\end{equation}
\begin{equation}
\hat{a}_{\mathbf{k},\lambda}\left|N_{\mathbf{k},\lambda}\right\rangle =\sqrt{N_{\mathbf{k},\lambda}}\left|N_{\mathbf{k},\lambda}-1\right\rangle .
\end{equation}

Therefore, the transition overlap in Eqn.~(\ref{eq:Fermi's-golden-rule})
separates into 
\begin{equation}
\left\langle i,n;N_{\mathbf{k},\lambda}\right|\hat{\mathbf{d}}\cdot\hat{\mathbf{E}}\left|f,m;N_{\mathbf{k},\lambda}\pm1\right\rangle =\left\langle i,n\left|\hat{\mathbf{d}}\right|f,m\right\rangle \otimes\left\langle N_{\mathbf{k},\lambda}\right|\hat{\mathbf{E}}\left|N_{\mathbf{k},\lambda}\pm1\right\rangle ,
\end{equation}
 and the full transition dipole matrix element becomes 
\begin{equation}
\left\langle i,n\left|\hat{\mathbf{d}}\right|f,m\right\rangle =\left\langle \chi_{i,n}\right|\left\langle i;\mathbf{R}\right|\hat{\mathbf{d}}\left|f;\mathbf{R}\right\rangle \left|\chi_{f,m}\right\rangle =\left\langle \chi_{i,n}\right|\mathbf{d}_{if}(\mathbf{R})\left|\chi_{f,m}\right\rangle ,
\end{equation}
 where 
\begin{equation}
\mathbf{d}_{if}(\mathbf{R})=\left\langle i;\mathbf{R}\right|\hat{\mathbf{d}}\left|f;\mathbf{R}\right\rangle \label{eq:optical-transition-dipole}
\end{equation}
 is the electronic transition dipole moment. In the Condon approximation
\cite{condonTheoryIntensityDistribution1926} its dependence on the
nuclear configuration is neglected, 
\begin{equation}
\mathbf{d}_{if}(\mathbf{R})\approx\mathbf{d}_{if}(\mathbf{R}_{0})=\mathbf{d}_{if},
\end{equation}
 such that the electronic and vibrational parts factorize as 
\begin{equation}
\left\langle i,n\left|\hat{\mathbf{d}}\right|f,m\right\rangle =\mathbf{d}_{if}\left\langle \chi_{i,n}\vert\chi_{f,m}\right\rangle .\label{eq:dipole-factorization}
\end{equation}
 The vibrational overlap appearing in Eqn.~(\ref{eq:dipole-factorization})
is the Franck-Condon amplitude responsible for redistributing the optical
intensity among the zero-phonon-line (ZPL) and phonon-assisted transitions \cite{alkauskasFirstPrinciplesCalculationsLuminescence2012,alkauskasFirstprinciplesTheoryLuminescence2014}.

\subsubsection*{Spontaneous emission process}

For spontaneous emission, Fermi's golden rule gives the transition
probability from an initial vibronic state $\left|i,n\right\rangle $,
with the photon field initially in the vacuum state $\left|0_{\mathbf{k},\lambda}\right\rangle $,
to $\left|f,m\right\rangle $ with one photon $\left|1_{\mathbf{k},\lambda}\right\rangle $,
through 
\begin{equation}
\Gamma^{\textnormal{em}}_{i,n\rightarrow f,m}=\frac{2\pi}{\hbar}\sum_{\mathbf{k},\lambda}\frac{\hbar\omega_{\mathbf{k}}}{2\epsilon_{0}n^{2}_{r}V}\left|\mathbf{d}_{if}\cdot\mathbf{e}_{\mathbf{k},\lambda}\right|^{2}\left|\left\langle \chi_{i,n}\vert\chi_{f,m}\right\rangle \right|^{2}\delta(E_{i,n}-E_{f,m}-\hbar\omega_{\mathbf{k}}).
\end{equation}
 Since the photon states are dense, the mode sum can be replaced by
an integral according to 
\begin{equation}
\sum_{\mathbf{k},\lambda}\longrightarrow\frac{V}{\left(2\pi\right)^{3}}\sum^{2}_{\lambda=1}\int\mathrm{d}^{3}k,
\end{equation}
 and, using the photon dispersion $\omega_{\mathbf{k}}=ck/n_{r}$,
the differential volume element in reciprocal space becomes 
\begin{equation}
\mathrm{d}^{3}k=\frac{n^{3}_{r}\omega^{2}}{c^{3}}\mathrm{d}\omega\mathrm{d}\Omega,
\end{equation}
 and the transition probability takes the form: 
\begin{equation}
\Gamma^{\textnormal{em}}_{i,n\rightarrow f,m}=\frac{n_{r}}{8\pi^{2}\epsilon_{0}c^{3}}\left|\left\langle \chi_{i,n}\vert\chi_{f,m}\right\rangle \right|^{2}\int^{\infty}_{0}\mathrm{d}\omega\,\omega^{3}\delta(E_{i,n}-E_{f,m}-\hbar\omega)\int\mathrm{d}\Omega\,\sum^{2}_{\lambda=1}\left|\mathbf{d}_{if}\cdot\mathbf{e}_{\mathbf{k},\lambda}\right|^{2}.
\end{equation}
 The transverse polarization sum satisfies 
\begin{equation}
\sum^{2}_{\lambda=1}\left|\mathbf{d}_{if}\cdot\mathbf{e}_{\mathbf{k},\lambda}\right|^{2}=\left|\mathbf{d}_{if}\right|^{2}-\left|\mathbf{d}_{if}\cdot\hat{\mathbf{k}}\right|^{2},
\end{equation}
 where $\hat{\mathbf{k}}$ has unit length and denotes the direction
of the wave vector $\mathbf{k}$. Thus, the angular integration over
the full solid angle gives 
\begin{equation}
\int\mathrm{d}\Omega\,\sum^{2}_{\lambda=1}\left|\mathbf{d}_{if}\cdot\mathbf{e}_{\mathbf{k},\lambda}\right|^{2}=\frac{8\pi}{3}\left|\mathbf{d}_{if}\right|^{2}.
\end{equation}
 Furthermore, the photon angular frequency integral simplifies to
\begin{equation}
\int^{\infty}_{0}\mathrm{d}\omega\,\omega^{3}\delta(E_{i,n}-E_{f,m}-\hbar\omega)=\frac{1}{\hbar}\left(\frac{E_{i,n}-E_{f,m}}{\hbar}\right)^{3},
\end{equation}
 altogether yielding a final expression for the spontaneous emission
transition probability in the form of 
\begin{equation}
\Gamma^{\textnormal{em}}_{i,n\rightarrow f,m}=\frac{n_{r}\left|\mathbf{d}_{if}\right|^{2}}{3\pi\epsilon_{0}\hbar c^{3}}\left(\frac{E_{i,n}-E_{f,m}}{\hbar}\right)^{3}\left|\left\langle \chi_{i,n}\vert\chi_{f,m}\right\rangle \right|^{2}.
\end{equation}

At finite temperature, the initial vibrational states are thermally
occupied according to 
\begin{equation}
p_{i,n}=\frac{\exp\left[-\beta\left(E_{i,n}-E_{i,0}\right)\right]}{\sum_{k}\exp\left[-\beta\left(E_{i,k}-E_{i,0}\right)\right]}=\frac{\exp\left[-\beta E_{i,n}\right]}{\sum_{k}\exp\left[-\beta E_{i,k}\right]},\qquad\beta=\frac{1}{k_{\textnormal{B}}T},\label{eq:phonon-temperature-occupation-probability}
\end{equation}
 where $k_{\textnormal{B}}$ is the Boltzmann constant and $T$ is
the thermodynamic temperature. It is therefore useful to separate
the purely vibrational redistribution of intensity from the photon
density-of-states factor by introducing a finite-temperature Lehmann-type
spectral representation for the vibronic transition in general as
\begin{equation}
A^{(\pm)}_{i\rightarrow f}(\hbar\omega)=\sum_{n,m}p_{i,n}\left|\left\langle \chi_{i,n}\vert\chi_{f,m}\right\rangle \right|^{2}\delta\left(\hbar\omega\pm\left[E_{i,n}-E_{f,m}\right]\right),\label{eq:spectral-function}
\end{equation}
 which for spontaneous emission takes the $A_{\textnormal{em}}(\hbar\omega)=A^{(-)}_{i\rightarrow f}(\hbar\omega)$ branch. 
 Finally, the energy-resolved spontaneous-emission
rate becomes 
\begin{equation}
W_{\textnormal{em}}(\hbar\omega)=\sum_{n,m}p_{i,n}\Gamma^{\textnormal{em}}_{i,n\rightarrow f,m}\delta\left(\hbar\omega-\left[E_{i,n}-E_{f,m}\right]\right)=\frac{n_{r}\left|\mathbf{d}_{if}\right|^{2}}{3\pi\epsilon_{0}\hbar c^{3}}\omega^{3}A_{\textnormal{em}}(\hbar\omega).
\end{equation}
 The characteristic $\omega^{3}$ dependence originates from the combination
of the photon density-of-states and the electric-field amplitude of
a single photon.

\subsubsection*{Absorption process}

For absorption, the incident photon occupies a specific mode $\left(\mathbf{k},\lambda\right)$,
since the incident laser beam has a specified direction, wavelength
and polarization. The initial photon field having $N_{\mathbf{k},\lambda}$
photons is in state $\left|N_{\mathbf{k},\lambda}\right\rangle $,
thus the annihilation operator in Eqn.~(\ref{eq:quantized-electric-field})
introduces an additional $N_{\mathbf{k},\lambda}$ factor. The photon-mode
sum reduces due to the specific incident mode $\left(\mathbf{k},\lambda\right)$
and Fermi's golden rule gives 
\begin{equation}
\Gamma^{\textnormal{abs}}_{i,n\rightarrow f,m}=\frac{\pi\omega_{\mathbf{k}}N_{\mathbf{k},\lambda}}{\epsilon_{0}n^{2}_{r}V}\left|\mathbf{d}_{if}\cdot\mathbf{e}_{\mathbf{k},\lambda}\right|^{2}\left|\left\langle \chi_{i,n}\vert\chi_{f,m}\right\rangle \right|^{2}\delta(E_{f,m}-E_{i,n}-\hbar\omega_{\mathbf{k}}).
\end{equation}
 For an isotropic ensemble (where emitters are randomly oriented),
\begin{equation}
\left\langle \left|\mathbf{d}_{if}\cdot\mathbf{e}_{\mathbf{k},\lambda}\right|^{2}\right\rangle =\frac{1}{3}\left|\mathbf{d}_{if}\right|^{2},
\end{equation}
 while the spectral representation of the vibronic transition takes
$A_{\textnormal{abs}}(\hbar\omega)=A^{(+)}_{i\rightarrow f}(\hbar\omega)$
to satisfy energy conservation. Thus, the energy-resolved absorption
rate becomes 
\begin{equation}
W_{\textnormal{abs}}(\hbar\omega)=\sum_{n,m}p_{i,n}\Gamma^{\textnormal{abs}}_{i,n\rightarrow f,m}\delta\left(\hbar\omega+\left[E_{i,n}-E_{f,m}\right]\right)=\frac{\pi N_{\mathbf{k},\lambda}\left|\mathbf{d}_{if}\right|^{2}}{3\epsilon_{0}n^{2}_{r}V}\omega A_{\textnormal{abs}}(\hbar\omega),
\end{equation}
 where the characteristic $\omega$ dependence now solely originates
from the electric-field amplitude of a single photon. However, this
rate depends on the incident photon flux $\Phi_{\mathbf{k},\lambda}$,
since for $N_{\mathbf{k},\lambda}$ photons in volume $V$ of an isotropic,
non-dispersive dielectric medium with refractive index $n_{r}$, the
photon flux is 
\begin{equation}
\Phi_{\mathbf{k},\lambda}=\frac{N_{\mathbf{k},\lambda}}{V}\frac{c}{n_{r}}=\Phi(\omega).
\end{equation}
 Hence, the absorption rate is 
\begin{equation}
W_{\textnormal{abs}}(\hbar\omega)=\Phi(\omega)\frac{\pi\left|\mathbf{d}_{if}\right|^{2}}{3\epsilon_{0}n_{r}c}\omega A_{\textnormal{abs}}(\hbar\omega),
\end{equation}
 from which the photon absorption cross-section is defined as 
\begin{equation}
\sigma_{\textnormal{abs}}(\hbar\omega)=\frac{W_{\textnormal{abs}}(\hbar\omega)}{\Phi(\omega)}=\frac{\pi\left|\mathbf{d}_{if}\right|^{2}}{3\epsilon_{0}n_{r}c}\omega A_{\textnormal{abs}}(\hbar\omega).
\end{equation}

Thus, the emission and absorption processes are governed by the same
underlying vibrational-overlap problem defined in Eqn.~(\ref{eq:spectral-function}).
However, in many experiments the absolute intensity scale is not known,
and instead normalized lineshapes are reported, conveniently defined
as 
\begin{equation}
L_{\textnormal{em}}(\hbar\omega)=\frac{\omega^{3}A_{\textnormal{em}}(\hbar\omega)}{\int\omega^{3}A_{\textnormal{em}}(\hbar\omega)\mathrm{d}(\hbar\omega)},
\end{equation}
 and 
\begin{equation}
L_{\textnormal{abs}}(\hbar\omega)=\frac{\omega A_{\textnormal{abs}}(\hbar\omega)}{\int\omega A_{\textnormal{abs}}(\hbar\omega)\mathrm{d}(\hbar\omega)}.
\end{equation}
 These expressions retain the physically relevant intensity redistribution
and required photon-energy dependence, while eliminating energy-independent
transition-dipole prefactors.

\subsection{Harmonic nuclear motion and the Huang-Rhys representation}

The remaining problem is the evaluation of the vibrational overlaps
appearing in Eqn.~(\ref{eq:spectral-function}). For an electronic
state $s$, the nuclear Hamiltonian can be expanded around the equilibrium
geometry $\mathbf{R}_{s,0}$. Within the harmonic approximation, the
nuclear Hamiltonian is 
\begin{equation}
\hat{H}_{s}=E_{\textnormal{min},s}-\sum_{a,\alpha}\frac{\hbar^{2}}{2M_{a}}\frac{\partial^{2}}{\partial R^{2}_{a,\alpha}}+\frac{1}{2}\sum_{a,\alpha;b\beta}u_{s;a,\alpha}\Phi_{s;a,\alpha;b,\beta}u_{s;b,\beta},\label{eq:harmonic-nuclear-hamiltonian-realspace}
\end{equation}
 where $M_{a}$ is the mass of atom $a$, 
\begin{equation}
u_{s;a,\alpha}=R_{a,\alpha}-R_{s,0;a,\alpha},
\end{equation}
 and 
\begin{equation}
\Phi_{s;a,\alpha;b,\beta}=\left.\frac{\partial^{2}E_{s}(\mathbf{R})}{\partial R_{a,\alpha}\partial R_{b,\beta}}\right|_{\mathbf{R}=\mathbf{R}_{s,0}}
\end{equation}
 is the force-constant matrix associated with electronic state $s$
around its equilibrium geometry, while $\alpha,\beta$ denote Cartesian
coordinate components. Introducing the mass-weighted displacement
vector 
\begin{equation}
\mathbf{x}_{s}=\mathbf{M}^{1/2}\left(\mathbf{R}-\mathbf{R}_{s,0}\right),\label{eq:mw-displacement}
\end{equation}
 where $\mathbf{M}$ denotes the nuclear mass matrix, which is diagonal
in the combined atomic and Cartesian indices, motivates the introduction
of the mass-weighted Hessian, or dynamical matrix, as 
\begin{equation}
\mathbf{D}_{s}=\mathbf{M}^{-1/2}\mathbf{\Phi}_{s}\mathbf{M}^{-1/2}.\label{eq:dynamical-matrix}
\end{equation}
 The dynamical matrix defines orthonormal eigenvectors through 
\begin{equation}
\mathbf{D}_{s}\mathbf{e}_{s,\nu}=\Omega^{2}_{s,\nu}\mathbf{e}_{s,\nu},\qquad\mathbf{e}^{\dagger}_{s,\nu}\mathbf{e}_{s,\mu}=\delta_{\nu\mu},\label{eq:dynamical-matrix-eigenproblem}
\end{equation}
 These eigenvectors define the normal coordinates of the atomic displacements as
\begin{equation}
Q_{s,\nu}=\mathbf{e}^{\dagger}_{s,\nu}\mathbf{x}_{s}.
\end{equation}
 The state-specific, harmonic nuclear Hamiltonian therefore separates
into independent harmonic oscillators, 
\begin{equation}
\hat{H}_{s}=E_{\textnormal{min},s}+\sum_{\nu}\left[\frac{\hat{P}^{2}_{s,\nu}}{2}+\frac{1}{2}\Omega^{2}_{s,\nu}\hat{Q}^{2}_{s,\nu}\right].\label{eq:harmonic-nuclear-hamiltonian-normalmode}
\end{equation}
 Naturally, the three rigid translational modes are excluded from
the set of physical vibrational modes for solids, while for molecules
an additional three modes are excluded describing the rigid body rotations.

In general, the normal modes of the two electronic states involved
in the transition need not coincide. Let $\mathbf{E}_{s}$ denote
the matrix whose columns are $\mathbf{e}_{s,\nu}$, thus it satisfies
$\mathbf{D}_{s}\mathbf{E}_{s}=\mathbf{\Omega}^{2}_{s}\mathbf{E}_{s}$,
where $\mathbf{\Omega}^{2}_{s}$ is diagonal, and define the mass-weighted
displacement between the two equilibrium geometries as 
\begin{equation}
\Delta\mathbf{x}=\mathbf{M}^{1/2}\left(\mathbf{R}_{f,0}-\mathbf{R}_{i,0}\right).
\end{equation}
 Since $\mathbf{x}_{f}=\mathbf{x}_{i}-\Delta\mathbf{x}$, the two
sets of normal coordinates are related by a Duschinsky transformation \cite{doktorovDynamicalSymmetryVibronic1977},
\begin{equation}
\mathbf{Q}_{f}=\mathbf{J}\mathbf{Q}_{i}+\mathbf{K},\label{eq:duschinsky}
\end{equation}
 where $\mathbf{J}=\mathbf{E}^{\dagger}_{f}\mathbf{E}_{i}$ is the
Duschinsky rotation matrix and $\mathbf{K}=-\mathbf{E}^{\dagger}_{f}\Delta\mathbf{x}$
is the displacement between the two minima expressed in the final-state
normal-mode basis. Eqn.~(\ref{eq:duschinsky}) describes the general
harmonic relation between the vibrational coordinates of the two electronic
states.

A full treatment must therefore account for both frequency changes
and Duschinsky mixing of modes, which requires a full multidimensional
Franck-Condon/Doktorov treatment \cite{doktorovDynamicalSymmetryVibronic1977},
which is outside the scope of the present work. In the Huang-Rhys
generating-function formulation adopted here, we make the parallel-mode
and equal-frequency approximations, setting 
\begin{equation}
J_{\nu\mu}=\delta_{\nu\mu},\qquad\Omega_{i,\nu}=\Omega_{f,\nu}\equiv\Omega_{\nu}.
\end{equation}
 The two potential-energy surfaces are then represented by identical
harmonic wells displaced along their common normal coordinates. Choosing
the minimum of the initial surface as the energy origin, the nuclear
motions are then described by the Hamiltonians, 
\begin{equation}
\hat{H}_{i}=\sum_{\nu}\left[\frac{\hat{P}^{2}_{\nu}}{2}+\frac{1}{2}\Omega^{2}_{\nu}\hat{Q}^{2}_{\nu}\right],
\end{equation}
 and 
\begin{equation}
\hat{H}_{f}=\Delta E+\sum_{\nu}\left[\frac{\hat{P}^{2}_{\nu}}{2}+\frac{1}{2}\Omega^{2}_{\nu}\left(\hat{Q}_{\nu}-\Delta Q_{\nu}\right)^{2}\right],
\end{equation}
 where $\Delta E=E_{\textnormal{min},f}-E_{\textnormal{min},i}$ is
the constant energy shift between the two minima, and the mode-resolved
mass-weighted displacement is 
\begin{equation}
\Delta Q_{\nu}=\mathbf{e}^{\dagger}_{\nu}\Delta\mathbf{x}=\sum_{a}\sqrt{M_{a}}\mathbf{e}^{\dagger}_{\nu,a}\Delta\mathbf{R}_{a},\label{eq:mode-resolved-mw-displacement}
\end{equation}
 where $\Delta\mathbf{R}_{a}=\mathbf{R}_{f,0;a}-\mathbf{R}_{i,0;a}$.

From Eqn.~(\ref{eq:harmonic-nuclear-hamiltonian-normalmode}), it
follows that the multidimensional vibrational wave functions factorize,
\begin{equation}
\left|\chi_{s,n}\right\rangle =\prod_{\nu}\left|n_{\nu}\right\rangle ,
\end{equation}
 where $\left|n_{\nu}\right\rangle $ represents the harmonic oscillator
along mode $\nu$ being in eigenstate $n_{\nu}$ with energy $\varepsilon=\hbar\Omega_{\nu}\left(n_{\nu}+\frac{1}{2}\right)$,
and therefore 
\begin{equation}
\left|\left\langle \chi_{i,n}\vert\chi_{f,m}\right\rangle \right|^{2}=\prod_{\nu}\left|\left\langle n_{\nu}\vert m_{\nu}\right\rangle \right|^{2}.\label{eq:overlap-factorization}
\end{equation}
 Katriel and Adam \cite{katrielUseSecondQuantization1970} showed that for two displaced, equal-frequency harmonic
oscillators the overlap between two states can be given by 
\begin{equation}
I_{n,m}(S)=\left\langle n\vert m\right\rangle =e^{S/2}\frac{\left(-\sqrt{S}\right)^{n-m}}{\sqrt{n!m!}}\sum^{\infty}_{l=\mathrm{max}\left(0,m-n\right)}\frac{\left(l+n\right)!}{l!\left(l+n-m\right)!}\left(-S\right)^{l},\label{eq:analytical-overlap-formula}
\end{equation}
 where the displacement of the two potentials is characterized entirely
by the partial Huang-Rhys factor, 
\begin{equation}
S=\frac{\Omega\left|\Delta Q\right|^{2}}{2\hbar},
\end{equation}
 originally introduced in the theory of vibronic transitions involving
 localized electronic states in crystals \cite{huangTheoryLightAbsorption1950}. The mode index $\nu$ was omitted here
  to improve readability.
When Eqn.~(\ref{eq:analytical-overlap-formula}) is restricted to
$n\geq m$, it can be brought to a closed form, since $\mathrm{max}\left(0,m-n\right)=0$
and thus 
\begin{equation}
\sum^{\infty}_{l=0}\frac{\left(l+n\right)!}{l!\left(l+n-m\right)!}\left(-S\right)^{l}=\sum^{\infty}_{l=0}m!\left(\begin{array}{c}
l+n\\
m
\end{array}\right)\frac{\left(-S\right)^{l}}{l!}.\label{eq:infinite-sum}
\end{equation}
 One can recognize that the right-hand side of Eqn.~(\ref{eq:infinite-sum})
is closely related to the generalized Laguerre polynomial as 
\begin{equation}
\sum^{\infty}_{l=0}\left(\begin{array}{c}
l+n\\
m
\end{array}\right)\frac{\left(x\right)^{l}}{l!}=e^{x}L^{(n-m)}_{m}(-x)
\end{equation}
 is the exponential generating function identity for shifted binomial
coefficients, where $L^{(\alpha)}_{n}(x)$ is the generalized Laguerre
polynomial. Therefore, the overlap has the 
\begin{equation}
I_{n,m}(S)=e^{-S/2}\left(-\sqrt{S}\right)^{n-m}\sqrt{\frac{m!}{n!}}L^{(n-m)}_{m}(S),\qquad n\geq m,\label{eq:overlap-restriced-closed-form}
\end{equation}
 closed form can be defined on a restricted domain. However, due to
the conjugate symmetry of the Hilbert space inner product,
\begin{equation}
\left\langle n\vert m\right\rangle =\left\langle m\vert n\right\rangle ^{*}\longrightarrow\quad\left|\left\langle n\vert m\right\rangle \right|^{2}=\left|\left\langle m\vert n\right\rangle \right|^{2},
\end{equation}
 Eqn.~(\ref{eq:overlap-restriced-closed-form}) can be used to define
\begin{equation}
\left|\left\langle n\vert m\right\rangle \right|^{2}=e^{-S}\left(S\right)^{n-m}\frac{m!}{n!}\left(L^{(n-m)}_{m}(S)\right)^{2}\label{eq:overlap-analytical-closed-form}
\end{equation}
 over the entire $n,m\geq0$ domain. 

Thus, the multidimensional vibrational overlap in Eqn.~(\ref{eq:spectral-function})
is expressed entirely by mode-specific partial Huang-Rhys factors,
\begin{equation}
S_{\nu}=\frac{\Omega_{\nu}\left|\Delta Q_{\nu}\right|^{2}}{2\hbar},\label{eq:partial-HR-factor}
\end{equation}
 which only depend on the vibrational-mode frequency from Eqn.~(\ref{eq:dynamical-matrix-eigenproblem}),
and the mode-resolved mass-weighted displacement from Eqn.~(\ref{eq:mode-resolved-mw-displacement}).

\subsection{Generating-function formulation}

Although the multidimensional vibrational overlap in Eqn.~(\ref{eq:spectral-function})
can be expressed analytically through Eqn.~(\ref{eq:overlap-analytical-closed-form}),
it still requires the evaluation of a double infinite sum over all
possible initial- and final-state vibrational occupations. The generating-function
framework makes it possible to evaluate this double infinite sum
and define a closed form for the spectral function of the vibronic
transition defined in Eqn.~(\ref{eq:spectral-function}). This approach has its 
origins in the generating-function theory of optical transitions in crystals 
\cite{laxFranckCondonPrincipleIts1952,kuboApplicationMethodGenerating1955}.

The central idea is to substitute a formal representation for the
$\delta$-distribution, through the inverse Fourier transform of the
tempered distribution $f(t)=1$ as 
\begin{equation}
\delta\left(x\right)=\frac{1}{2\pi}\int^{+\infty}_{-\infty}f(t)e^{-ixt}\,\mathrm{d}t=\frac{1}{2\pi}\int^{+\infty}_{-\infty}e^{-ixt}\,\mathrm{d}t.\label{eq:formal-diracdelta-distribution}
\end{equation}
 Since the argument of the $\delta$-distribution in Eqn.~(\ref{eq:spectral-function})
has units of energy, it is better to use $t/\hbar$ as the Fourier-conjugate
variable to $x$ instead of just $t$. Thus, after substitution, the
spectral function becomes 
\begin{equation}
A^{(\pm)}_{i\rightarrow f}(\hbar\omega)=\frac{1}{2\pi\hbar}\int^{+\infty}_{-\infty}\sum_{n,m}p_{i,n}\left|\left\langle \chi_{i,n}\vert\chi_{f,m}\right\rangle \right|^{2}\mathrm{\exp}\left\{ -i\left(\hbar\omega\pm\left[E_{i,n}-E_{f,m}\right]\right)\frac{t}{\hbar}\right\} \,\mathrm{d}t,
\end{equation}
 where the integration and summation order was exchanged, and where
the extra $\hbar^{-1}$ factor originates from the $t\rightarrow t/\hbar$
variable change when compared to Eqn.~(\ref{eq:formal-diracdelta-distribution}).
Using 
\begin{equation}
\frac{E_{i,n}-E_{f,m}}{\hbar}=\frac{E_{i,0}-E_{f,0}}{\hbar}-\sum_{\nu}\Omega_{\nu}\left(m_{\nu}-n_{\nu}\right)=\Delta\omega_{0}-\sum_{\nu}\Omega_{\nu}\left(m_{\nu}-n_{\nu}\right),
\end{equation}
 the spectral function can be written as 
\begin{equation}
A^{(\pm)}_{i\rightarrow f}(\hbar\omega)=\frac{1}{2\pi\hbar}\int^{+\infty}_{-\infty}e^{-i\left(\omega\pm\Delta\omega_{0}\right)t}G^{(\pm)}(t)\,\mathrm{d}t,
\end{equation}
 where 
\begin{equation}
G^{(\pm)}(t)=\sum_{n,m}p_{i,n}\left|\left\langle \chi_{i,n}\vert\chi_{f,m}\right\rangle \right|^{2}\mathrm{\exp}\left\{ \pm i\sum_{\nu}\Omega_{\nu}\left(m_{\nu}-n_{\nu}\right)t\right\} 
\end{equation}
 is the generating function of the vibronic transition. From Eqn.~(\ref{eq:overlap-factorization})
it follows that the generating function also factorizes, 
\begin{equation}
G^{(\pm)}(t)=\prod_{\nu}G^{(\pm)}_{\nu}(t),\qquad G^{(\pm)}_{\nu}(t)=\sum_{n_{\nu},m_{\nu}}p_{i,n_{\nu}}\left|\left\langle n_{\nu}\vert m_{\nu}\right\rangle \right|^{2}e^{\pm i\left(m_{\nu}-n_{\nu}\right)\Omega_{\nu}t}.
\end{equation}
The contribution of a single mode $\nu$ to the full generating function, obtained using Eqn.~(\ref{eq:overlap-analytical-closed-form}), is 
\begin{equation}
\tilde{G}^{(\pm)}(t)=\sum_{n,m}p_{i,n}e^{-S}\left(S\right)^{n-m}\frac{m!}{n!}\left(L^{(n-m)}_{m}(S)\right)^{2}e^{\pm i\left(m-n\right)\Omega t},
\end{equation}
 where the mode index was dropped again to improve readability. From
Eqn.~(\ref{eq:phonon-temperature-occupation-probability}) and the
harmonic approximation it follows that 
\begin{equation}
p_{i,n}=\left(1-e^{-\beta\hbar\Omega}\right)e^{-n\beta\hbar\Omega},
\end{equation}
 and thus the generating function factor of a single mode reads as
\begin{equation}
\tilde{G}^{(\pm)}(t)=\left(1-e^{-\beta\hbar\Omega}\right)e^{-S}\sum^{\infty}_{n=0}e^{-n\beta\hbar\Omega}\frac{1}{n!}\left(\frac{S}{e^{\pm i\Omega t}}\right)^{n}\sum^{\infty}_{m=0}\frac{m!}{S^{m}}\left(L^{(n-m)}_{m}(S)\right)^{2}e^{\pm im\Omega t}.\label{eq:generating-function-explicit-summation}
\end{equation}
 One can recognize that in the equation above (Eqn.~(\ref{eq:generating-function-explicit-summation})),
\begin{equation}
\sum^{\infty}_{m=0}\frac{m!}{S^{m}}\left(L^{(n-m)}_{m}(S)\right)^{2}e^{\pm im\Omega t}=\sum^{\infty}_{m=0}m!\left(\frac{e^{\pm i\Omega t}}{S}\right)^{m}\left(L^{(n-m)}_{m}(S)\right)^{2}
\end{equation}
 is of the form 
\begin{equation}
\sum^{\infty}_{m=0}m!\left(x\right)^{m}L^{(n-m)}_{m}(y)L^{(n-m)}_{m}(z)=n!x^{n}e^{xyz}L^{(0)}_{n}\left(-\frac{\left(1-xy\right)\left(1-xz\right)}{x}\right),
\end{equation}
 which is a Mehler-type bilinear generating function, coming from
the $L^{(n-m)}_{m}(S)=\left[u^{m}\right]e^{-Su}\left(1+u\right)^{n}$
coefficient-extraction identity\footnote{Here the $\left[x^{m}\right]f(x)$ notation is used for the coefficient
of $x^{m}$ in the power-series expansion of $f(x)$.}. With this, 
\begin{equation}
\sum^{\infty}_{m=0}m!\left(\frac{e^{\pm i\Omega t}}{S}\right)^{m}\left(L^{(n-m)}_{m}(S)\right)^{2}=n!\left(\frac{e^{\pm i\Omega t}}{S}\right)^{n}\mathrm{\exp}\left\{ Se^{\pm i\Omega t}\right\} L^{(0)}_{n}\left(-\frac{S\left(1-e^{\pm i\Omega t}\right)^{2}}{e^{\pm i\Omega t}}\right),
\end{equation}
 and Eqn.~(\ref{eq:generating-function-explicit-summation}) simplifies
to 
\begin{equation}
\tilde{G}^{(\pm)}(t)=\left(1-e^{-\beta\hbar\Omega}\right)e^{-S}\mathrm{\exp}\left\{ Se^{\pm i\Omega t}\right\} \sum^{\infty}_{n=0}e^{-n\beta\hbar\Omega}L^{(0)}_{n}\left(-\frac{S\left(1-e^{\pm i\Omega t}\right)^{2}}{e^{\pm i\Omega t}}\right).
\end{equation}
 Similarly, the summation over $n$ can also be evaluated since it is
in the form of the standard generating function for generalized Laguerre
polynomials \cite{hardySummationSeriesPolynomials1932,hilleLaguerresSeries1926a},
\begin{equation}
\sum^{\infty}_{n=0}t^{n}L^{(\alpha)}_{n}(x)=\frac{\mathrm{\exp}\left\{ -\frac{tx}{1-t}\right\} }{\left(1-t\right)^{\alpha+1}},
\end{equation}
 yielding 
\begin{equation}
\tilde{G}^{(\pm)}(t)=\left(1-e^{-\beta\hbar\Omega}\right)e^{-S}\mathrm{\exp}\left\{ Se^{\pm i\Omega t}\right\} \frac{1}{1-e^{-\beta\hbar\Omega}}\mathrm{\exp}\left\{ \frac{e^{-\beta\hbar\Omega}}{1-e^{-\beta\hbar\Omega}}\frac{S\left(1-e^{\pm i\Omega t}\right)^{2}}{e^{\pm i\Omega t}}\right\} .
\end{equation}
 Introducing the thermal phonon occupation number, 
\begin{equation}
\overline{n}_{\nu}=\frac{1}{\exp\left(\beta\hbar\Omega_{\nu}\right)-1},
\end{equation}
 the full generating function of the vibronic transition becomes 
\begin{equation}
G^{(\pm)}(t)=\prod_{\nu}G^{(\pm)}_{\nu}(t)=\prod_{\nu}\mathrm{\exp}\left\{ S_{\nu}\left[\left(\bar{n}_{\nu}+1\right)e^{\pm i\Omega_{\nu}t}+\bar{n}_{\nu}e^{\mp i\Omega_{\nu}t}-\left(2\bar{n}_{\nu}+1\right)\right]\right\} .\label{eq:remove}
\end{equation}

At the absolute zero-temperature limit ($\bar{n}_{\nu}\rightarrow0$)
the generating function simplifies to 
\begin{equation}
G^{(\pm)}(t)=\mathrm{\exp}\left\{ \sum_{\nu}S_{\nu}\left[e^{\pm i\Omega_{\nu}t}-1\right]\right\} ,
\end{equation}
 where the summation over individual phonon modes can be replaced by
an integration, such that 
\begin{equation}
G^{(\pm)}(t)=\mathrm{\exp}\left\{ \int^{+\infty}_{0}S\left(\hbar\omega\right)\left[e^{\pm i\omega t}-1\right]\mathrm{d}\left(\hbar\omega\right)\right\} ,
\end{equation}
 and the introduced function $S\left(\hbar\omega\right)$ is the Huang-Rhys
spectral-density, describing how strongly phonon modes contribute
to the transition at specific energies. For individual phonon modes,
it is defined as a sum of $\delta$-distributions weighted by partial
Huang-Rhys factors,
\begin{equation}
S(\hbar\omega)=\sum_{\nu}S_{\nu}\delta\left(\hbar\omega-\hbar\Omega_{\nu}\right).\label{eq:final-generator-form-start}
\end{equation}
 Additionally, at the zero-temperature limit the $m$-phonon progression
follows a Poisson distribution, 
\begin{equation}
G^{(\pm)}(t)=e^{-S_{\textnormal{tot.}}}\mathrm{\exp}\left\{ \sum_{\nu}S_{\nu}e^{\pm i\Omega_{\nu}t}\right\} =e^{-S_{\textnormal{tot.}}}\mathrm{\exp}S(\pm t)=e^{-S_{\textnormal{tot.}}}\sum^{\infty}_{m=0}\frac{\left[S(\pm t)\right]^{m}}{m!},
\end{equation}
 where $S_{\textnormal{tot.}}=\sum_{\nu}S_{\nu}$ is the total Huang-Rhys
factor, and therefore the zero-phonon contribution carries the weight
\begin{equation}
w_{\textnormal{ZPL}}(T=0)=e^{-S_{\textnormal{tot.}}}.
\end{equation}
 Therefore, at zero-temperature, the zero-phonon contribution carries the weight $w_{\textnormal{ZPL}}=e^{-S_{\textnormal{tot.}}}$, corresponding to the Debye-Waller factor within the displaced-harmonic-oscillator approximation \cite{huangTheoryLightAbsorption1950,alkauskasFirstprinciplesTheoryLuminescence2014}.
 Furthermore, the introduced Fourier transform of Eqn.~(\ref{eq:final-generator-form-start})
is 
\begin{equation}
S(t)=\sum_{\nu}S_{\nu}e^{i\Omega_{\nu}t},\label{eq:fourier-transformed-HR-spectral-density}
\end{equation}
 defined through 
\begin{equation}
S(\hbar\omega)=\frac{1}{2\pi\hbar}\int^{+\infty}_{-\infty}S(t)e^{-i\omega t}\mathrm{d}t.
\end{equation}

At this point, we define alternate quantities, 
\begin{equation}
\bar{S}\left(t\right)=S\left(t\right)-S\left(0\right)=S\left(t\right)-S_{\textnormal{tot.}}=\sum_{\nu}S_{\nu}\left(e^{i\Omega_{\nu}t}-1\right),
\end{equation}
 and 
\begin{equation}
\bar{S}_{\text{th.}}\left(T;t\right)=\sum_{\nu}\bar{n}_{\nu}S_{\nu}\left(e^{i\Omega_{\nu}t}-1\right),
\end{equation}
to rewrite the generating function as 
\begin{equation}
G^{\left(\pm\right)}\left(t\right)=\exp\left\{ \bar{S}\left(\pm t\right)+\bar{S}_{\text{th.}}\left(T;t\right)+\bar{S}_{\text{th.}}\left(T;-t\right)\right\} .\label{eq:final-generator-form-end}
\end{equation}
 The sign convention defined in Eqn.~(\ref{eq:spectral-function}),
which accounts for energy conservation in the emission and absorption
processes, makes it clear that by default, optical transitions can
only be assisted by phonon creation, defined by $\bar{S}\left(\pm t\right)$.
Since temperature-dependent processes appear in the generating function
with both time evolution signs, defined by $\bar{S}_{\text{th.}}\left(T;t\right)$,
both phonon creation and annihilation channels activate with non-zero
temperature symmetrically. Consequently, the thermal activation of
the annihilation process introduces an anti-Stokes contribution to
the spectral-function, manifested in the intensity redistribution
above and below the ZPL energy for emission and absorption, respectively.

Equations~(\ref{eq:final-generator-form-start})-(\ref{eq:final-generator-form-end})
establish the central connection between the one-phonon Huang-Rhys
spectral-density $S(\hbar\omega)$ and the complete multiphonon optical
lineshape $L_{\textnormal{em/abs}}(\hbar\omega)$. The ideal harmonic
expressions contain $\delta$-distributions and therefore infinitely
sharp spectral structures. To compare with experiment, additional
broadening can be introduced after the intrinsic vibronic spectral
function $A^{(\pm)}_{i\rightarrow f}(\hbar\omega)$ has been constructed.
A Lorentzian convolution with half-width $\gamma$, for example, is
equivalent in the time domain to multiplication by 
\begin{equation}
e^{-\gamma\left|t\right|/\hbar},
\end{equation}
 whereas convolution with a Gaussian of standard deviation $\sigma$
corresponds to 
\begin{equation}
e^{-\sigma^{2}t^{2}/\left(2\hbar^{2}\right)}.
\end{equation}
 These broadenings can be used to represent homogeneous and inhomogeneous
contributions, respectively, and should be distinguished from the
electron-phonon broadening already contained in the generating function.

\subsection{Force representation of the Huang-Rhys factors}

The preceding formulation expresses the electron-phonon coupling through
the difference between the equilibrium geometries of the two electronic
states. Within the harmonic approximation, the same information can
be written exactly in terms of the force generated by the electronic
transition.

For a transition from an initial state $i$ to a final state
$f$, define 
\begin{equation}
\mathbf{\Delta R}=\mathbf{R}_{f,0}-\mathbf{R}_{i,0},
\end{equation}
 to be the displacement in equilibrium atomic positions as defined above. Immediately after a vertical electronic
transition, the nuclei remain at $\mathbf{R}_{i,0}$, while the electronic
state has changed to the final electronic state $f$. Since $\mathbf{R}_{i,0}$
is an equilibrium geometry of the initial-state, 
\begin{equation}
\mathbf{F}_{i}\left(\mathbf{R}_{i,0}\right)=0,
\end{equation}
 where the subscript on the force denotes that it is evaluated for
the $i$ initial-state electronic structure. The transition-induced
force is therefore 
\begin{equation}
\mathbf{F}^{\textnormal{tr}}=\mathbf{F}_{f}\left(\mathbf{R}_{i,0}\right)-\mathbf{F}_{i}\left(\mathbf{R}_{i,0}\right)=\mathbf{F}_{f}\left(\mathbf{R}_{i,0}\right),
\end{equation}
 which is related to the total energy by 
\begin{equation}
\mathbf{F}^{\textnormal{tr}}=-\left.\nabla_{\mathbf{R}}E_{f}(\mathbf{R})\right|_{\mathbf{R}=\mathbf{R}_{i,0}},\label{eq:general-transition-force}
\end{equation}
 where $E_{f}(\mathbf{R})$ is the total energy for the $f$ final-state
electronic structure in atomic geometry $\mathbf{R}$. Expanding this
final-state potential around its minimum in the harmonic potential
gives 
\begin{equation}
E_{f}(\mathbf{R})\simeq E_{\textnormal{min},f}+\frac{1}{2}\left(\mathbf{R}-\mathbf{R}_{f,0}\right)^{T}\mathbf{\Phi}_{f}\left(\mathbf{R}-\mathbf{R}_{f,0}\right),
\end{equation}
 previously introduced in Eqn.~(\ref{eq:harmonic-nuclear-hamiltonian-realspace}),
where $\mathbf{\Phi}_{f}$ denotes the force-constant matrix for the $f$ final-state
electronic structure around its equilibrium geometry $\mathbf{R}_{f,0}$.
Thus, in the harmonic approximation the force felt by the atoms due
to being out of equilibrium is simply 
\begin{equation}
\mathbf{F}_{f}\left(\mathbf{R}\right)=-\mathbf{\Phi}_{f}\left(\mathbf{R}-\mathbf{R}_{f,0}\right).\label{eq:harmonic-finalstate-force}
\end{equation}
 Consequently, the transition-induced force, obtained by substituting
Eqn.~(\ref{eq:harmonic-finalstate-force}) into Eqn.~(\ref{eq:general-transition-force}),
is 
\begin{equation}
\mathbf{F}^{\textnormal{tr}}=\mathbf{\Phi}_{f}\mathbf{\Delta R}.
\end{equation}
 Equivalently, the transition-induced atomic relaxation is 
\begin{equation}
\mathbf{\Delta R}=\mathbf{\Phi}^{-1}_{f}\mathbf{F}^{\textnormal{tr}},\label{eq:displacement-from-force}
\end{equation}
 where $\mathbf{\Phi}^{-1}_{f}$ is the pseudo-inverse of the force-constant
matrix understood on the subspace where zero-energy translational
(and rotational) modes are excluded. 

Eqn.~(\ref{eq:displacement-from-force}) separates the perturbation
caused by the electronic transition from the elastic response of the
lattice. Even when the transition-induced force is spatially concentrated
around the defect, the resulting displacement field need not be similarly
confined because it is obtained by acting on the force source with
the static lattice Green function $\mathbf{\Phi}^{-1}$. In reciprocal
space, the static lattice Green function contains a strongly enhanced
$q^{-2}$ long-wavelength contribution for acoustic modes. Consequently,
a finite periodically repeated defect supercell can represent the
local source relatively well while its relaxed displacement field
still contains an appreciable response to the periodic array of defect
images \cite{varvennePointDefectModeling2013,clouetElasticModelingPointdefects2018}.

This formulation provides a convenient way to reconstruct the transition-force
source from fully relaxed structures. Since the harmonic approximation
is not exact in general, rather than directly obtaining transition-induced
forces from ab initio calculations by constraining the geometry and
the electronic structure, we find it more appropriate to compute an
effective transition-induced force $\tilde{\mathbf{F}}^{\textnormal{tr}}$
from an ab initio calculated $\mathbf{\Delta\tilde{R}}$ displacement
and an appropriate force-constant matrix $\tilde{\mathbf{\Phi}}_{f}$.
Using the relaxed displacement and the harmonic force-constant matrix 
avoids treating anharmonic contributions in directly calculated vertical 
forces as purely harmonic. 

As discussed above, it is more convenient to define the force-source
in mass-weighted coordinates, where 
\begin{equation}
\Delta\mathbf{x}=\mathbf{M}^{1/2}\mathbf{\Delta R},
\end{equation}
 and consequently 
\begin{equation}
\mathbf{f}^{\textnormal{tr}}=\mathbf{D}_{f}\mathbf{\Delta x}=\mathbf{M}^{-1/2}\mathbf{F}^{\textnormal{tr}},
\end{equation}
 where $\mathbf{D}_{f}$ and $\mathbf{M}$ are the dynamical matrix
and nuclear mass matrix, respectively, defined previously in Eqn.~(\ref{eq:mw-displacement})
and (\ref{eq:dynamical-matrix}). Expanding the mass-normalized force-source
in terms of the dynamical matrix eigenvectors $\mathbf{e}_{f;\nu}$,
defined in Eqn.~(\ref{eq:dynamical-matrix-eigenproblem}), gives
\begin{equation}
\mathbf{f}^{\textnormal{tr}}=\mathbf{D}_{f}\mathbf{\Delta x}=\mathbf{D}_{f}\sum_{\nu}\Delta Q_{\nu}\mathbf{e}_{f;\nu}=\sum_{\nu}\Omega^{2}_{f,\nu}\Delta Q_{\nu}\mathbf{e}_{f;\nu},
\end{equation}
 and projecting it onto mode $\nu$ yields 
\begin{equation}
\mathbf{e}^{\dagger}_{f;\nu}\mathbf{f}^{\textnormal{tr}}=\Omega^{2}_{f,\nu}\Delta Q_{\nu}.
\end{equation}
 Therefore, 
\begin{equation}
\Delta Q_{\nu}=\frac{\mathbf{e}^{\dagger}_{f;\nu}\mathbf{f}^{\textnormal{tr}}}{\Omega^{2}_{f,\nu}},
\end{equation}
 and substituting into Eqn.~(\ref{eq:partial-HR-factor}) gives a
force representation for the partial Huang-Rhys factors, 
\begin{equation}
S_{\nu}=\frac{\left|\mathbf{e}^{\dagger}_{f;\nu}\mathbf{f}^{\textnormal{tr}}\right|^{2}}{2\hbar\Omega^{3}_{f,\nu}}.\label{eq:partial-HR-factor-force_Supp}
\end{equation}
 Eqns.~(\ref{eq:partial-HR-factor}) and (\ref{eq:partial-HR-factor-force_Supp})
are mathematically equivalent within the assumed harmonic approximation.
The partial Huang-Rhys factors can be expressed by both the relaxed
atomic displacement and the transition-induced force-source that generates
that displacement.

\subsection{Interpolation of the vibrational continuum}

A direct first-principles calculation in a defect supercell containing
$N$ atoms provides only $3N-3$ non-translational vibrational modes.
The corresponding Huang-Rhys spectral-density is therefore represented
by a finite collection of discrete frequencies. This sparse sampling
is particularly restrictive for long-wavelength acoustic modes and
for fine structure associated with critical points of the phonon dispersion.
The force formulation above makes it possible to separate the local
transition source from the subsequent vibrational response and to
sample the latter on a substantially denser set of modes.

\begin{figure}

\begin{centering}
\includegraphics{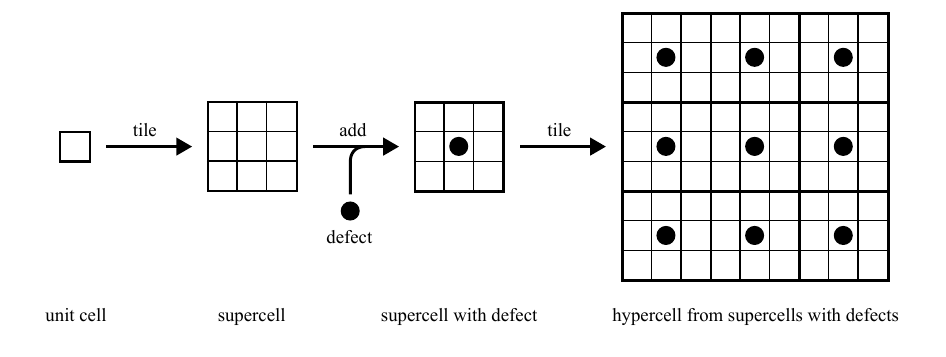}\caption{\textbf{Schematic construction of the hypercell.} A pristine unit
cell is tiled to form a pristine supercell, into which the defect
center under investigation is introduced. This defect-containing supercell
is then used as the repeating unit of a larger hypercell.}\label{fig:hypercell-schematic}
\par\end{centering}
\end{figure}

We achieve this by regarding the defect-containing first-principles
supercell as the elementary periodic cell. The resulting hypercell
(supercell of supercells) is obtained by repeating the defect-containing
supercell 
\begin{equation}
M_{1}\times M_{2}\times M_{3}
\end{equation}
 times, containing altogether 
\begin{equation}
N_{q}=M_{1}M_{2}M_{3}
\end{equation}
 copies, as shown schematically in Fig.~\ref{fig:hypercell-schematic}.
Equivalently, the vibrational modes of this hypercell can be obtained
from an $M_{1}\times M_{2}\times M_{3}$ mesh of wave vectors in the
Brillouin-zone of the defect supercell, without explicitly constructing
and diagonalizing the full real-space hypercell dynamical matrix \cite{giannozziInitioCalculationPhonon1991,gonzeDynamicalMatricesBorn1997b}.

Let $A$ and $B$ label two periodic images of the defect-containing
supercell within this hypercell. Then the force-constant matrix element
coupling ion $a$ in image $A$ to ion $b$ in image $B$ is 
\begin{equation}
\Phi_{a,\alpha;b,\beta}(A,B)=\left.\frac{\partial^{2}E(\mathbf{R})}{\partial R_{a,\alpha}(A)\,\partial R_{b,\beta}(B)}\right|_{\mathbf{R}=\mathbf{R}_{0}},
\end{equation}
where $R_{a,\alpha}(A)$ denotes the Cartesian coordinate $\alpha$
of ion $a$ in supercell image $A$. Due to translational invariance,
originating from the periodic construction, 
\begin{equation}
\Phi_{a,\alpha;b,\beta}(A,B)=\Phi_{a,\alpha;b,\beta}(A+C,B+C),
\end{equation}
 meaning that the interaction between different supercell images 
depends only on their relative translation
\begin{equation}
L=B-A,
\end{equation}
since the real-space force-constants satisfy
\begin{equation}
\Phi_{a,\alpha;b,\beta}(A,B)=\Phi_{a,\alpha;b,\beta}(0,L).
\end{equation}
 This translational structure allows the finite real-space force-constant
kernel to be transformed to reciprocal space according to 
\begin{equation}
\Phi_{a,\alpha;b,\beta}(\mathbf{q})=\sum_{L}\Phi_{a,\alpha;b,\beta}(0,L)e^{i\mathbf{q}\cdot\left[\mathbf{R}_{b}(L)-\mathbf{R}_{a}(0)\right]}.
\end{equation}
 After mass-weighting, one obtains the $\mathbf{q}$-dependent dynamical
matrix 
\begin{equation}
D_{a,\alpha;b,\beta}(\mathbf{q})=\frac{\Phi_{a,\alpha;b,\beta}(\mathbf{q})}{\sqrt{M_{a}M_{b}}},
\end{equation}
 from which the interpolated phonon frequencies $\Omega_{\nu}(\mathbf{q})$
and eigenvectors $e_{\nu;a,\alpha}(\mathbf{q})$ can be obtained by
diagonalizing this matrix independently at each commensurate $\mathbf{q}$
wave vector, 
\begin{equation}
\sum_{b,\beta}D_{a,\alpha;b,\beta}(\mathbf{q})e_{\nu;b,\beta}(\mathbf{q})=\Omega^{2}_{\nu}(\mathbf{q})e_{\nu;a,\alpha}(\mathbf{q}).
\end{equation}

The eigenvectors $e_{\nu;a,\alpha}(\mathbf{q})$ describe the phonon
displacement pattern inside one defect-containing supercell. The corresponding
normalized, real-space Bloch eigenvector of the full hypercell is
\begin{equation}
E_{\nu;L,a,\alpha}(\mathbf{q})=\frac{1}{\sqrt{N_{q}}}e^{i\mathbf{q}\cdot\mathbf{R}_{a}(L)}e_{\nu;a,\alpha}(\mathbf{q}),
\end{equation}
 where the $N^{-1/2}_{q}$ prefactor ensures the normalization to
unity across the repeated supercells. 

The electronic transition, however, is intended to represent the transition
of a single isolated defect rather than a coherent transition of all
periodically repeated defects in the hypercell. Therefore the mass-normalized
transition-induced force-source is defined as a localized source acting
within only one reference supercell, 
\begin{equation}
\mathbf{f}^{\textnormal{tr}}_{L,a}=\mathbf{f}^{\textnormal{tr}}_{a}\delta_{L,0}.\label{eq:local-force-source}
\end{equation}
 Applying the force-based representation of the partial Huang-Rhys
factor in Eqn.~(\ref{eq:partial-HR-factor-force_Supp}) to the hypercell
modes gives 
\begin{equation}
S_{\mathbf{q},\nu}=\frac{\left|\sum_{L,a}\mathbf{E}^{\dagger}_{\nu;L,a}(\mathbf{q})\:\mathbf{f}^{\textnormal{tr}}_{L;a}\right|^{2}}{2\hbar\Omega^{3}_{\nu}(\mathbf{q})},\label{eq:partal-HR-factor-hypercell-phonon-mode}
\end{equation}
 while also imposing the local source requirement in Eqn.~(\ref{eq:local-force-source})
yields 
\begin{equation}
S_{\mathbf{q},\nu}=\frac{1}{2N_{q}\hbar\Omega^{3}_{\nu}(\mathbf{q})}\left|\sum_{a}\mathbf{e}^{\dagger}_{\nu;a}(\mathbf{q})\:\mathbf{f}^{\textnormal{tr}}_{a}e^{-i\mathbf{q}\mathbf{R}_{a}(0)}\right|^{2}.\label{eq:partial-HR-factors-hypercell}
\end{equation}

The exact phase appearing in Eqn.~(\ref{eq:partial-HR-factors-hypercell})
depends on the convention adopted for the Bloch eigenvectors. Physical
Huang-Rhys factors are independent of this convention, provided the
force-constants, eigenvectors, and force-source are treated consistently. 

The $N^{-1}_{q}$ factor in Eqn.~(\ref{eq:partial-HR-factors-hypercell})
originates directly from the normalization of the extended hypercell
phonons. It ensures that upon increasing the hypercell size, the total
Huang-Rhys factor remains a properly normalized, convergent quantity.
The interpolated Huang-Rhys spectral-density is therefore 
\begin{equation}
S\left(\hbar\omega\right)=\sum_{\mathbf{q},\nu}S_{\mathbf{q},\nu}\,\delta\left(\hbar\omega-\hbar\Omega_{\nu}(\mathbf{q})\right),\label{eq:interpolated-HR-spectral-density_Supp}
\end{equation}
 and 
\begin{equation}
S_{\textnormal{tot.}}=\sum_{\mathbf{q},\nu}S_{\mathbf{q},\nu}
\end{equation}
 approaches a well-defined continuum limit as the $\mathbf{q}$-point
sampling is increased.

The interpolated Huang-Rhys spectral-density obtained from Eqn.~(\ref{eq:interpolated-HR-spectral-density_Supp})
is inserted without further modification into the finite-temperature
generating function of Eqn.~(\ref{eq:fourier-transformed-HR-spectral-density})-(\ref{eq:final-generator-form-end}).
Thus, the full calculation forms a closed sequence, \begin{equation}
	\begin{tikzcd}[column sep=large, row sep=normal]
		\Delta\mathbf R
		    \arrow[r]
		&
		\mathbf F^{\textnormal{tr}}
		    \arrow[r]
		&
		S_{\mathbf q\nu}
		    \arrow[dl]
		    \arrow[dr]
		&
		\mathbf D(\mathbf q)
			\arrow[l]
		\\
		&
		S(\hbar\omega)
		    \arrow[rr, dashed]
		&&
		G(t)
		    \arrow[r]
		&
		A_{\mathrm{em/abs}}(\hbar\omega)
		    \arrow[r]
		&
		L_{\mathrm{em/abs}}(\hbar\omega)
	\end{tikzcd}
\end{equation} linking the first-principles structural response of the defect to
the observable optical lineshape.

For finite numerical sampling, the $\delta$-distributions, for example
in Eqn.~(\ref{eq:interpolated-HR-spectral-density_Supp}), are generally
replaced by normalized kernels, 
\begin{equation}
\delta\left(\hbar\omega-\hbar\Omega_{\nu}(\mathbf{q})\right)\rightarrow K_{\eta}\left(\hbar\omega-\hbar\Omega_{\nu}(\mathbf{q})\right),
\end{equation}
 with a small width $\eta$. Any smearing applied to the Huang-Rhys
spectral-density serves only as a continuous representation for the
discretely sampled continuum and should be distinguished from the
homogeneous and inhomogeneous optical broadenings applied to the final
lineshape.

\subsection{Scope of the theoretical framework}

The theory above provides a quantum-mechanical finite-temperature
description of multiphonon optical sidebands while retaining a tractable
first-principles formulation. The introduced approximations facilitate
analytical progress at each step. The Born-Oppenheimer approximation
separates electronic and nuclear motion, while the Condon approximation
neglects the variation of the electronic transition dipole with nuclear
coordinates. Nuclear motion is treated harmonically around the two
equilibrium structures. The parallel-mode approximation neglects Duschinsky
rotation between the two sets of normal modes, and the equal-frequency
approximation replaces their generally different curvatures by a common
vibrational spectrum. Subject to these assumptions, the generating-function
expression includes arbitrarily high multiphonon order and Bose-Einstein
thermal populations exactly. 

While these approximations make it possible to capture the driving
effect in a transition, they also impose limitations on applicability.
Effects outside this framework include anharmonic nuclear motion,
phonon lifetimes arising from anharmonicity, non-Condon or Herzberg-Teller
contributions to the optical matrix element, Duschinsky mixing, and
explicit vibronic coupling associated with electronically degenerate
states such as Jahn-Teller systems \cite{doktorovDynamicalSymmetryVibronic1977,razinkovasVibrationalVibronicStructure2021}. 

Similarly, the force-based interpolation, introduced above and in
the main text, does not by itself remove errors already present in
the underlying electronic-state geometries or force constants. Its
purpose is instead to reconstruct the transition-induced source within
the same harmonic theory and couple this source to a substantially
denser representation of the vibrational spectrum. When the defect
supercell is sufficiently converged to represent the local transition-induced
force and local force constants, increasing the hypercell size then
predominantly improves the sampling of the long-wavelength and fine-structure
components of the phonon continuum rather than altering the local
electron-phonon coupling itself.

\section{Spatial extent of transition-induced forces \label{sec:spatial-extent-of-forces}}

\subsection{Local, analytic contribution \label{sec:local-force-contribution}}

An optical transition between two electronic states modifies the electronic charge distribution and consequently the forces acting on the surrounding ions. Within the Born-Oppenheimer approximation, the transition-induced force on ion $a$ is defined as the difference between the forces associated with the two electronic potential-energy surfaces,
\begin{equation}
    \mathbf{F}^\textnormal{tr}_a=-\frac{\partial}{\partial\mathbf{R}_a}\left(E_e-E_g\right)=\mathbf{F}^\textnormal{e}_a-\mathbf{F}^\textnormal{g}_a,
\end{equation}
where $E_g$ and $E_e$ are the ground- and excited-state electronic energies respectively, evaluated at a common atomic configuration.

The spatial extent of these forces is closely connected to the electronic redistribution accompanying an optical transition. Introducing the difference between the electronic densities,
\begin{equation}
    \Delta n(\mathbf{r})=n_e(\mathbf{r})-n_g(\mathbf{r}),
\end{equation}
the Hellmann-Feynman theorem \cite{feynmanForcesMolecules1939} relates the transition force to the interaction between $\Delta n$ and the electron-ion potential. For a local ionic potential, this contribution can be written schematically as 
\begin{equation}
    F^\textnormal{tr}_{a,i}=-\int d^3r\,\Delta n(\mathbf{r})\frac{\partial v_a\left(\mathbf{r}-\mathbf{R}_a\right)}{\partial R_{a,i}},
\end{equation}
with additional terms when nonlocal pseudopotentials are employed.

For optical transitions involving deep defect states in an insulating host, the change in electronic occupation is primarily associated with orbitals localized around the defect. These orbitals are localized states within the band gap, and the corresponding wavefunctions generally possess exponentially decaying spatial envelopes \cite{kohnWannierFunctionsSimple1973}.

Accordingly, the local bonding contribution to the transition force may be represented by an asymptotic envelope of the form
\begin{equation}
    \left|\mathbf{F}^\textnormal{an}(\mathbf{r})\right|\lesssim Ar^{-n}e^{-r/\xi},
\end{equation}
where $A$ characterizes the force amplitude and $\xi$ is an effective force-localization length. The algebraic prefactor follows from the specific exponential character of the localized orbitals. Importantly, $\xi$ need not coincide with the localization length of the defect orbitals, since the force also depends on the electronic screening response and the spatial structure of the electron-ion interaction.

The exponential localization of the underlying defect orbitals thus provides a physical basis for a short-range transition-force contribution. However, it does not establish that the complete transition force is exponentially localized. Although the electronic redistribution associated with the change in defect occupation is confined to the vicinity of the defect, its electrostatic multipole moments can generate electric fields extending over much larger distances. These fields may couple to the ionic degrees of freedom of the surrounding crystal, producing additional contributions with algebraic spatial decay. It is therefore useful to distinguish the localized analytic part of the force from possible long-range electrostatic terms,
\begin{equation}
    \mathbf{F}^\textnormal{tr}=\mathbf{F}^\textnormal{an}+\mathbf{F}^\textnormal{NA},
\end{equation}
where the superscripts indicate the contribution that are analytic and nonanalytic when  expressed in reciprocal space. The nonanalytic terms arise due to the macroscopic electrostatic response of the host.

Since the rigid translation of the crystal cannot change the electronic transition energy, the net transition force must vanish,
\begin{equation}
    \sum_a F^\textnormal{tr}_{a,i} = 0.
\end{equation}
Both the analytic and nonanalytic long-range force components can be defined consistently such that this translational force-neutrality constraint is respected by each contribution.

The Fourier transform of the localized force distribution is
\begin{equation}
    \tilde{F}^\textnormal{an}_i(\mathbf{q})=\mathcal{F}[F^\textnormal{an}_i(\mathbf{r})](\mathbf{q})=\mathcal{F}\left[\sum_a F^\textnormal{an}_{a,i}\delta(\mathbf{r}-\mathbf{R}_a)\right](\mathbf{q})=\int d^3r\sum_a F^\textnormal{an}_{a,i}\delta(\mathbf{r}-\mathbf{R}_a)e^{-i\mathbf{q}\mathbf{r}}=\sum_a F^\textnormal{an}_{a,i}e^{-i\mathbf{q}\mathbf{R}_a}.
\end{equation}
Due to the exponential localization of this force contribution, all its spatial moments are finite, and its Fourier transform is analytic in the neighborhood of $\mathbf{q}=0$. Expanding in powers of the wave vector gives
\begin{equation}
    \tilde{F}^\textnormal{an}_i(\mathbf{q})=\sum_a F^\textnormal{an}_{a,i} -iq_j\sum_a F^\textnormal{an}_{a,i}R_{a,j}+\mathcal{O}(q^2)=-iq_jP^\textnormal{tr}_{i,j}+\mathcal{O}(q^2),
\end{equation}
where the $P^\textnormal{tr}_{i,j}$ is the transition-induced elastic-force dipole. On the right-hand side, Einstein-notation is used, so repeated indices are summed. The same notation will be used below. This term provides the leading continuum representation of the localized transition-force source, which is used in the elastic-dipole description of point defects \cite{clouetElasticModelingPointdefects2018}.

\subsection{Long-range electrostatic and electromechanical coupling \label{sec:long-range-force-contribution}}

The bare electric charge-density difference associated with an optical transition is
\begin{equation}
    \Delta\rho(\mathbf{r})=\rho_e(\mathbf{r})-\rho_g(\mathbf{r}),
\end{equation}
with ionic positions held fixed. The long-range electrostatic field is determined by the multipole moments of this source. Its Fourier transform may be expanded as
\begin{equation}
    \Delta\tilde{\rho}(\mathbf{q})=\mathcal{F}[\Delta\rho(\mathbf{r})](\mathbf{q})=Q-iq_ip_i-\frac{1}{2}q_iq_jQ_{ij}+\mathcal{O}(q^3),
\end{equation}
where $Q$, $p_i$ and $Q_{i,j}$ denote the difference in the total charges, the dipole moments and the quadrupole moments respectively. For a charge-conserving optical transition
\begin{equation}
    Q=\int d^3r\,\Delta\rho(\mathbf{r})=0,
\end{equation}
and the leading electrostatic contribution is therefore dipolar, unless the electric dipole difference also vanishes by symmetry. The difference of the permanent electric dipoles $\mathbf{p}$ of the two electronic states is a separate quantity and should not be confused with the optical transition-dipole matrix element $\mathbf{d}_{if}$  introduced in Eqn.~(\ref{eq:optical-transition-dipole}), or with the elastic-force dipole $P^\textnormal{tr}_{i,j}$ introduced above.

In a homogeneous insulating medium with dielectric tensor $\mathbf{\epsilon}$, the screened electrostatic potential $\phi$ satisfies
\begin{equation}
    \nabla\cdot\left(\epsilon_0\mathbf{\epsilon}\nabla\phi\right)=-\Delta\rho,
\end{equation}
which, expressed in reciprocal space, yields
\begin{equation}
    \tilde{\phi}(\mathbf{q})=\mathcal{F}[\phi(\mathbf{r})](\mathbf{q})=\frac{\Delta\tilde{\rho}(\mathbf{q})}{\epsilon_0q_i\epsilon_{ij}q_j}.
\end{equation}
The electric field in reciprocal space is simply $\tilde{E}_i(\mathbf{q})=-iq_i\tilde{\phi}(\mathbf{q})$. Thus, for a dipolar source, $\Delta\tilde{\rho}(\mathbf{q})\sim q$, the electric field contains a contribution of the form
\begin{equation}
    \tilde{E}_i(\mathbf{q})\propto\frac{q_i(\mathbf{q\cdot\mathbf{p}})}{q_j\epsilon_{jk}q_k},
\end{equation}
which remains finite in the $q\rightarrow0$ limit, and its value generally depends on the direction of $\mathbf{q}$. Therefore the electric field is nonanalytic at the Brillouin-zone center, despite originating from a spatially localized charge distribution $\Delta\rho$.

In real space, the familiar electrostatic multipole hierarchy follows. If the first nonvanishing electric source multipole has order $l$, with $l = 0, 1, 2$ corresponding to a monopole, dipole, and quadrupole, respectively, the electric field decays as
\begin{equation}
    |\mathbf{E}^{(l)}(\mathbf{r})|\sim r^{-(l+2)}.
\end{equation}

Therefore for a charge-preserving transition with nonzero electric dipole difference, the resulting far field decreases as $r^{-3}$. This electric field may act directly on individual ions through their Born effective charges. Thus in a general insulating crystal, the electrostatic forces can retain the $r^{-3}$ decay associated with a dipolar source. However, this contribution does not couple to long-wavelength acoustic phonon modes, since the Born effective charges within a unit cell satisfy an acoustic sum rule. This makes the response to a spatially uniform field cancel for the rigid-translation channel. The corresponding forces may instead act on relative ionic coordinates, associated with optical or internal phonon modes \cite{gonzeDynamicalMatricesBorn1997b}.

The electrostatic force that does couple to a slowly varying acoustic phonon mode is more naturally described through the electromechanical response of the host. Let $\mathbf{u}(\mathbf{r})$ denote a smooth displacement field and
\begin{equation}
    \eta_{jk}=\frac{1}{2}\left(\partial_ju_k+\partial_ku_j\right)
\end{equation}
its strain tensor. The polarization induced by this deformation admits a long-wavelength expansion,
\begin{equation}
    P^\textnormal{ind}_i=e_{ijk}\eta_{jk}+\mu_{ijkl}\partial_l\eta_{jk}+\dots,
\end{equation}
where $e_{ijk}$ and $\mu_{ijkl}$ are the piezoelectric and flexoelectric tensors, respectively \cite{tagantsevPiezoelectricityFlexoelectricityCrystalline1986,hongFirstprinciplesTheoryCalculation2013,stengelFlexoelectricityDensityfunctionalPerturbation2013}. The piezoelectric term describes polarization induced by strain, whereas the flexoelectric term describes polarization induced by a strain gradient. Since strain is even under spatial inversion, whereas polarization is odd, linear bulk piezoelectricity is forbidden in centrosymmetric crystals. Flexoelectricity, on the other hand, couples polarization to strain gradients, which are odd under spatial inversion, and is therefore symmetry-allowed even in centrosymmetric insulators.

The interaction between the transition-induced electric field and this polarization is described by
\begin{equation}
    U_\textnormal{int}=-\int d^3r\,E^\textnormal{tr}_i(\mathbf{r})P^\textnormal{ind}_i(\mathbf{r}).
\end{equation}
Varying this energy with respect to the displacement and integrating by parts shows that the acoustic force associated with piezoelectricity contains one spatial derivative of the electric field, while the flexoelectric contribution contains two. These contributions may be schematically expressed as
\begin{equation}
    F^\textnormal{pz}(\mathbf{r})\sim e\nabla E^\textnormal{tr}(\mathbf{r}), \qquad F^\textnormal{fx}(\mathbf{r})\sim \mu\nabla\nabla E^\textnormal{tr}(\mathbf{r}),
\end{equation}
where tensor contractions have been suppressed for clarity.

Combining these relations with the multipole decay of the electric field obtained above yields the possible asymptotic force envelopes of 
\begin{equation}
    F^\textnormal{pz}_{(l)}(r)\sim r^{-(l+3)}, \qquad F^\textnormal{fx}_{(l)}(r)\sim r^{-(l+4)}. \label{eq:piezo-flexo-force-tail}
\end{equation}
Thus, for a charge-preserving transition with nonzero electric dipole difference, the leading piezoelectric and flexoelectric forces can decay as $r^{-4}$ and $r^{-5}$, respectively. If the host is centrosymmetric, the piezoelectric contribution vanishes by symmetry and only the flexoelectric contribution is retained with $r^{-5}$ spatial decay in leading order.

The same qualification appears in reciprocal space.  Since $\Delta\tilde{\rho}^{(l)}(\mathbf{q})\sim q^l$ for multipole order $l$, the electric field scales as $q^{l-1}$. The additional derivatives in the electromechanical response then yield
\begin{equation}
    \tilde{F}^\textnormal{pz}_{(l)}(\mathbf{q})\sim q^l, \qquad \tilde{F}^\textnormal{fx}_{(l)}(\mathbf{q})\sim q^{l+1}.
\end{equation}

\subsection{Spatial crossover between localized and long-range force contributions}

The preceding discussion suggests that the transition-force field can contain two quantitatively different spatial contributions: an exponentially localized force associated with the change in chemical bonding, and an algebraically decaying electrostatic force mediated by the host response.

Writing the asymptotic magnitudes as
\begin{equation}
    F^\textnormal{an}(r)\sim Ar^{-n}e^{-r/\xi}, \qquad F^\textnormal{NA}(r)\sim B_k(\hat{\mathbf{r}})r^{-k},
\end{equation}
their ratio becomes
\begin{equation}
    \frac{|F^\textnormal{an}(r)|}{|F^\textnormal{NA}(r)|}\sim\frac{A}{|B_k(\hat{\mathbf{r}})|}r^{k-n}e^{-r/\xi},
\end{equation}
and therefore the relative importance of these terms depends on more than their decay exponents. Close to the defect, the transition modifies the occupation of localized orbitals and the bonding forces on neighboring atoms. If the corresponding local force amplitude is sufficiently large compared with the electromechanical coefficient, the analytic contribution can dominate over the defect environment and an intermediate range of distances.

However, at sufficiently large distances any nonzero algebraic contribution ultimately decays more slowly than the exponential term. An outer crossover then occurs at a distance determined by the force-localization length and the relative amplitudes of the two mechanisms. This crossover cannot be inferred from the asymptotic powers alone, and the continuum expressions should not be extrapolated into the atomistic defect core.

It is therefore possible for the localized contribution to account for most of the physically relevant transition-force weight, even though the complete force has an algebraic far-field tail. Beyond the crossover, the nonanalytic component may be mathematically dominant while its absolute magnitude is already negligibly small. The spatial dominance of the force and its contribution to the vibrational spectrum must consequently be assessed separately.

\subsection{Contributions to the Huang-Rhys spectral density in the acoustic region}

\begin{table}[t]
\centering

\begingroup
\renewcommand{\arraystretch}{1.5}
\setlength{\tabcolsep}{12pt}

\begin{tabular}{
    >{\centering\arraybackslash}m{3cm}
    >{\centering\arraybackslash}m{3.2cm}
    >{\centering\arraybackslash}m{1.5cm}
    >{\centering\arraybackslash}m{2.5cm}
    >{\centering\arraybackslash}m{3.3cm}
}
\toprule

\multirow[t]{2}{*}{\parbox[t]{3cm}{\centering Source}}
& \multirow[t]{2}{*}{\parbox[t]{3.2cm}{\centering Coupling mechanism}}
& \multicolumn{2}{c}{Force scaling}
& \multirow[t]{2}{*}{\parbox[t]{3.3cm}{\centering
    Acoustic $S(\hbar\omega)$ scaling\\
    ($\hbar\omega\rightarrow 0^+$)}} \\

& & \multicolumn{1}{>{\centering\arraybackslash}p{1.5cm}}{real-space}
& \multicolumn{1}{>{\centering\arraybackslash}p{2.5cm}}{
    reciprocal-space\newline ($q\rightarrow 0$)}
& \\

\midrule

Localized electronic redistribution
& Short-range bonding response
& $r^{-n}e^{-r/\xi}$
& $q$
& $\hbar\omega$ \\

Electric dipole
& Piezoelectricity
& $r^{-4}$
& $q$
& $\hbar\omega$ \\

Electric dipole
& Flexoelectricity
& $r^{-5}$
& $q^2$
& $(\hbar\omega)^3$ \\

Electric quadrupole
& Piezoelectricity
& $r^{-5}$
& $q^2$
& $(\hbar\omega)^3$ \\

Electric quadrupole
& Flexoelectricity
& $r^{-6}$
& $q^3$
& $(\hbar\omega)^5$ \\

\bottomrule
\end{tabular}

\endgroup

\caption{Leading spatial decay and acoustic spectral-density
scaling of localized and electrostatic transition-force
contributions in three-dimensional insulating crystals.}
\label{tab:force-scaling}

\end{table}

The transition force couples to a phonon mode through its projection onto the corresponding mass-normalized eigenvector. From Eqn.~(\ref{eq:partal-HR-factor-hypercell-phonon-mode}) and (\ref{eq:interpolated-HR-spectral-density_Supp}), the Huang-Rhys spectral density is expressed as
\begin{equation}
    S\left(\hbar\omega\right)=\sum_{\mathbf{q},\nu}\frac{\left|G_{\mathbf{q},\nu}\right|^{2}}{2\hbar\Omega^{3}_{\nu}(\mathbf{q})}\,\delta\left(\hbar\omega-\hbar\Omega_{\nu}(\mathbf{q})\right),
\end{equation}
where $\Omega_{\nu}(\mathbf{q})$ is the phonon frequency, and $G_{\mathbf{q},\nu}$ denotes the projected force amplitude with appropriate mode normalization.

For a three-dimensional crystal, the acoustic phonon branches satisfy
\begin{equation}
    \Omega_{\nu}(\mathbf{q})=c_\nu(\hat{\mathbf{q}})q + \mathcal{O}(q^2)
\end{equation}
in the $q\rightarrow0$ limit, where $\hat{\mathbf{q}}$ is a unit vector indicating directional dependence for $c_\nu(\hat{\mathbf{q}})$. Supposing that the transition force has the leading reciprocal space behavior
\begin{equation}
    \tilde{F}^\textnormal{tr}(\mathbf{q})\sim q^\beta
\end{equation}
implies
\begin{equation}
    G_{\mathbf{q},\nu}^{(\beta)}\sim q^\beta A_\nu(\hat{\mathbf{q}}),
\end{equation}
where $A_\nu(\hat{\mathbf{q}})$ is a finite angular coefficient. The contribution to the spectral density then scales as
\begin{equation}
    S^{(\beta)}\left(\hbar\omega\right)\propto\int q^2dq\,\frac{q^{2\beta}}{q^3}\,\delta\left(\hbar\omega-\hbar cq\right)\propto\left(\hbar\omega\right)^{2\beta-1},\quad(\hbar\omega\rightarrow0^+).
\end{equation}

Therefore, the linear acoustic spectral density is a natural consequence of a localized, force-balanced transition source with nonzero elastic-force dipole. For a nonvanishing piezoelectric tensor, the associated force is also proportional to $q$, contributing at the same linear order as the analytic force. The flexoelectric force, by contrast, begins at order $q^2$, yielding a spectral contribution proportional to $(\hbar\omega)^3$, as summarized in Table~\ref{tab:force-scaling}. Higher order electric multipoles and electromechanical couplings lead to correspondingly higher powers of $q$.

The piezoelectric and flexoelectric coupling mechanisms underlying these scalings are described by established electromechanical response theories \cite{tagantsevPiezoelectricityFlexoelectricityCrystalline1986,hongFirstprinciplesTheoryCalculation2013,stengelFlexoelectricityDensityfunctionalPerturbation2013}. The specific low-frequency powers of the Huang-Rhys spectral density then follow from combining those response laws with the transition-force multipole expansion and acoustic-mode projection derived above.

\section{Computational scaling of lineshape methods}

The computational requirements of the different optical lineshape methodologies can be divided into two stages: the construction of the underlying model from first-principles calculations and the subsequent evaluation of the vibrational or dynamical quantities required to obtain the optical spectrum. The relative importance of these stages varies considerably between methods. In particular, approaches that construct an effective large system from a smaller reference calculation separate the expensive electronic-structure problem from the vibrational sampling problem.

We denote the number of ions in the target system by $N$ and that in the smaller reference system by $n$, where $n\ll N$. For machine-learning-based approaches, $N_{\mathrm{train}}$ denotes the number of training configurations, while $N_t$ and $N_{\mathrm{traj}}$ are the number of MD time steps per trajectory and the number of independent trajectories, respectively. The scaling estimates summarized in Fig.~1 of the main text concern the dominant operations and assume fixed numerical convergence criteria unless stated otherwise. System-dependent prefactors, symmetry reductions, and parallelization efficiencies are omitted.

\subsection{Electronic-structure calculations and numerical preliminaries}

The computational cost of a DFT calculation depends on the exchange–correlation functional, basis representation, and numerical algorithms. For conventional plane-wave implementations employing semilocal functionals such as PBE, a commonly adopted asymptotic estimate is $\mathcal{O}(N^3)$. Although evaluating the semilocal exchange–correlation functional itself is comparatively inexpensive, the optimization and orthogonalization of the Kohn–Sham orbitals, together with the diagonalization of the electronic subspace, can introduce cubic scaling. Efficient iterative algorithms may exhibit lower effective exponents over practically relevant system sizes \cite{kresseEfficientIterativeSchemes1996}.

Hybrid functionals, such as HSE, introduce an additional nonlocal exact-exchange contribution. Its evaluation requires interactions between occupied electronic orbitals and is therefore considerably more expensive than semilocal exchange. In a conventional FFT-based plane-wave implementation, applying the exact-exchange operator to all occupied states involves approximately $\mathcal{O}(N_{\mathrm{occ}}^2N_G\log N_G)$ operations, where $N_{\mathrm{occ}}$ and $N_G$ denote the numbers of occupied orbitals and plane-wave grid points, respectively \cite{barnesImprovedTreatmentExact2017}. For increasing supercell size at fixed density and $k$-point sampling, both quantities scale approximately linearly with the number of ions, giving $\mathcal{O}(N^3\log N)$. Other exchange algorithms can exhibit different asymptotic behavior, and modern compressed-exchange techniques can substantially reduce computational costs \cite{linAdaptivelyCompressedExchange2016}. Consequently, the cubic DFT scaling adopted in Fig.~1 of the main text should be understood as a representative approximation for conventional electronic-structure calculations, rather than a universal scaling law for either PBE or HSE.

Within the finite-displacement approach, determining the complete interatomic force-constant (IFC) matrix of an $N$-ion system requires displacing a number of independent ionic degrees of freedom proportional to $N$ and evaluating the resulting forces \cite{togoFirstPrinciplesPhonon2015}. Neglecting symmetry reductions, this gives

\begin{equation}
C_{\mathrm{IFC}}(N)
\sim \mathcal{O}\!\left(N C_{\mathrm{DFT}}(N)\right).
\end{equation}

For the representative cubic DFT scaling, the IFC calculation therefore requires $\mathcal{O}(N^4)$ operations. The fourth-order scaling shown in the figure originates from the product of $\mathcal{O}(N)$ displaced configurations and the assumed $\mathcal{O}(N^3)$ cost of each DFT calculation. If the electronic-structure algorithm has a different effective scaling, the overall IFC cost changes accordingly.

Once the force constants are available, the phonon frequencies and eigenvectors are obtained from the dynamical matrix. For $N$ ions, this is a $3N\times3N$ matrix. Conventional dense diagonalization requires $\mathcal{O}(N^3)$ operations and $\mathcal{O}(N^2)$ memory when the complete eigenbasis is calculated \cite{andersonLAPACKUsersGuide1999}. The distinction between constructing the force constants and solving the resulting eigenvalue problem is central to the computational comparison below.

\subsection{Direct supercell calculations}

In the direct method, equilibrium geometries and vibrational properties are determined explicitly for the complete $N$-ion defect supercell. The IFC matrix is obtained using first-principles force calculations, and the resulting dynamical matrix is diagonalized to evaluate the phonon frequencies, eigenvectors, and partial Huang–Rhys factors.

The finite-displacement calculation is the dominant computational contribution, requiring $\mathcal{O}(N^4)$ operations under the assumed cubic DFT scaling. By comparison, dense diagonalization of the $3N\times3N$ dynamical matrix requires only $\mathcal{O}(N^3)$ operations. Projection of the equilibrium displacement onto all phonon eigenvectors involves at most $\mathcal{O}(N^2)$ additional operations and is therefore subdominant.

Thus, increasing the target system size requires repeating the most expensive electronic-structure calculations on the enlarged supercell. This makes the direct approach particularly demanding when the large cells required to resolve long-wavelength acoustic modes are considered. The use of hybrid functionals further increases the computational expense, often making a complete first-principles IFC calculation impractical.

The principal memory contributions arise from storing the electronic orbitals, the IFC matrix, and the phonon eigenvectors. At fixed plane-wave cutoff and $k$-point sampling, the electronic wavefunctions require approximately $\mathcal{O}(N^2)$ storage, as do the dense IFC matrix and complete phonon eigenbasis. Consequently, the characteristic memory scaling is $\mathcal{O}(N^2)$, although the actual memory requirement and prefactors depend on the electronic-structure implementation.

\subsection{Embedding method}

The embedding approach separates the calculation of local defect properties from the construction of the extended vibrational system. Following Alkauskas \textit{et al.} \cite{alkauskasFirstprinciplesTheoryLuminescence2014}, the force constants describing the defect region are obtained from a smaller first-principles calculation and combined with those of the pristine host to construct an effective dynamical matrix for a much larger system. The transition-induced forces can likewise be obtained from the smaller defect supercell and embedded into the extended system.

The first-principles calculation now serves as a prerequisite for constructing the effective model. If the reference defect supercell contains $n$ ions, its IFC calculation requires approximately $\mathcal{O}(n^4)$ operations under the same DFT assumptions as above. Importantly, this cost does not increase when the target system size $N$ is subsequently enlarged while keeping the reference model fixed.

The dominant system-size-dependent cost is therefore transferred to the vibrational calculation. The effective dynamical matrix has dimensions $3N\times3N$, and conventional dense diagonalization requires $\mathcal{O}(N^3)$ operations. This is the representative scaling indicated in Fig.~1 of the main text.

However, the short-ranged nature of the IFCs in many covalent materials makes the extended dynamical matrix sparse. If the interaction range is fixed, the number of nonzero matrix elements grows only as $\mathcal{O}(N)$, allowing the matrix itself to be constructed and stored with approximately linear complexity. Iterative sparse eigensolvers, such as Lanczos- or Arnoldi-based methods, exploit this structure through repeated sparse matrix–vector multiplications \cite{hernandezSLEPcScalableFlexible2005}. Each such multiplication requires $\mathcal{O}(N)$ operations, although the total cost also depends on the number of requested eigenpairs, convergence iterations, and orthogonalization operations.

Consequently, sparse diagonalization does not have a universal scaling exponent. For a fixed number of selected eigenpairs and favorable convergence, nearly linear scaling can be achieved. Obtaining the complete eigenbasis is substantially more expensive, however, and conventional full-spectrum calculations can approach cubic scaling. Moreover, explicitly storing all eigenvectors necessarily requires $\mathcal{O}(N^2)$ memory.

If only the force-projected vibrational spectral density is required, the complete eigenbasis can in principle be avoided. Krylov-subspace techniques or polynomial spectral expansions, such as the kernel polynomial method, can evaluate spectral quantities using repeated sparse matrix–vector operations \cite{weisseKernelPolynomialMethod2006}. For a fixed number of expansion moments, such approaches can achieve approximately linear computational and memory scaling. These methods nevertheless introduce additional convergence requirements associated with the spectral approximation and need not reproduce individual phonon eigenvectors.

Thus, the embedding approach can reduce the first-principles cost substantially, while the subsequent computational requirements depend strongly on the numerical method employed to solve the extended vibrational problem. The sparse IFC matrix requires $\mathcal{O}(N)$ memory, whereas retaining the complete phonon eigenbasis increases the requirement to $\mathcal{O}(N^2)$. The reference DFT calculation contributes an additional $\mathcal{O}(n^2)$ memory requirement.

\subsection{Phonon interpolation (present work)}

The interpolation method presented in this work similarly separates the first-principles determination of local defect properties from the vibrational calculation of the extended system. Equilibrium displacements, transition-induced forces, and IFCs are obtained from a reference defect supercell containing $n$ ions. The corresponding first-principles IFC calculation requires approximately $\mathcal{O}(n^4)$ operations and constitutes the model-construction stage.

The essential computational distinction is that the vibrational problem is solved in reciprocal space. The real-space IFCs of the reference supercell are Fourier transformed to construct a wave-vector-dependent dynamical matrix, following the general principle of phonon interpolation \cite{giannozziInitioCalculationPhonon1991}. For each wave vector $\mathbf q$, the dynamical matrix has fixed dimensions $3n\times3n$, irrespective of the size of the final effective system.

An effective hypercell containing $N$ ions corresponds to $N/n$ sampled wave vectors. Dense diagonalization of each $3n\times3n$ matrix requires $\mathcal{O}(n^3)$ operations, giving the total cost

\begin{equation}
C_{\mathrm{interp}}
=
\mathcal{O}\left(\frac{N}{n}n^3\right)
=
\mathcal{O}(Nn^2).
\end{equation}

The calculation of the partial Huang–Rhys factors from the phonon eigenvectors and localized transition force involves a lower-order contribution of $\mathcal{O}(Nn)$ when the force projections are evaluated directly for all modes.

For a fixed reference supercell, the interpolation cost therefore increases linearly with the target system size, while retaining the complete phonon spectrum at every sampled wave vector. Unlike real-space embedding, the method does not require diagonalizing a single dynamical matrix whose dimension increases with $N$. Instead, the calculation consists of independent, fixed-size eigenvalue problems, allowing straightforward parallelization over wave vectors.

This decomposition also substantially reduces memory requirements. The reference IFC representation, the dynamical matrix at one wave vector, and its complete eigenbasis require approximately $\mathcal{O}(n^2)$ memory. If individual wave vectors are processed sequentially and their contributions accumulated directly into the spectral density, this memory requirement is independent of $N$, apart from the chosen spectral accumulation grid. Storing all interpolated eigenvectors simultaneously would instead require $\mathcal{O}(Nn)$ memory, but is not necessary for evaluating the spectral density.

The computational advantage therefore becomes especially pronounced when the reference supercell is sufficiently large to converge the local defect perturbation, while much larger effective systems are needed to resolve the vibrational continuum. Increasing $N$ improves phonon sampling without increasing the dimension of the individual diagonalization problems. The accuracy of the underlying force constants and localized transition-force representation, however, remains controlled by the reference supercell size $n$.

\subsection{Machine-learning phonon calculations}

In the machine-learning approach of Turiansky \textit{et al.} \cite{turianskyMachineLearningPhonon2026}, the expensive direct evaluation of the IFC matrix is replaced by force constants obtained from a machine-learned interatomic potential. A pretrained model is fine-tuned using first-principles energies and forces, including configurations generated during conventional DFT structural relaxations.

If the training dataset contains $N_{\mathrm{train}}$ configurations of $n$ ions, the generation of the reference data requires approximately $\mathcal{O}(N_{\mathrm{train}}n^3)$ operations under the adopted DFT scaling. This constitutes the model-construction cost indicated in Fig.~1 of the main text. The optimization of the machine-learning model introduces an additional computational contribution whose scaling depends on the network architecture, training procedure, and number of optimization epochs. This training cost is not included explicitly in the figure.

Once the potential has been trained, the energy derivatives required to construct the IFC matrix can be evaluated without additional first-principles calculations. For local machine-learned potentials with a fixed interaction range and model architecture, such as MACE \cite{batatiaMACEHigherOrder2022}, the number of relevant atomic interactions grows linearly with system size. An implementation exploiting this locality can therefore evaluate and assemble the sparse IFC representation with approximately $\mathcal{O}(N)$ complexity. The actual cost can be higher if the Hessian is constructed through an algorithm that does not exploit its sparsity.

Although this greatly reduces the cost of obtaining the force constants, it does not by itself eliminate the extended phonon eigenvalue problem. The resulting dynamical matrix still has dimensions $3N\times3N$, and conventional complete diagonalization requires $\mathcal{O}(N^3)$ operations, as indicated in the figure. Sparse iterative solvers and direct spectral methods may reduce this cost under the conditions discussed for embedding.

The memory requirements likewise depend on the numerical representation. The DFT training data require approximately $\mathcal{O}(N_{\mathrm{train}}n)$ storage for atomic configurations, energies, and forces, in addition to the model parameters. For fixed-range interactions, the extended sparse IFC matrix can require only $\mathcal{O}(N)$ storage, while explicitly retaining the complete phonon eigenbasis requires $\mathcal{O}(N^2)$ memory. The reference electronic-structure calculations additionally retain their $\mathcal{O}(n^2)$ memory requirements.

Thus, machine learning primarily removes the expensive first-principles force-constant bottleneck. The computational scaling of the subsequent vibrational calculation remains determined by the chosen eigensolver and spectral representation.

\subsection{Molecular-dynamics autocorrelation method}

The MD–ACF approach of Linderälv \textit{et al.} \cite{linderalvOpticalLineShapes2025b} differs fundamentally from the preceding methods by avoiding explicit phonon diagonalization. Instead, the optical lineshape is reconstructed from the autocorrelation function of electronic energy-gap fluctuations sampled along classical MD trajectories.

The first-principles calculations again constitute a prerequisite for model construction. In this case, training configurations are required to represent the potential-energy surfaces of the relevant electronic states. For $N_{\mathrm{train}}$ configurations containing $n$ ions, the DFT data-generation cost is approximately $\mathcal{O}(N_{\mathrm{train}}n^3)$, with an additional model-dependent contribution from training the machine-learned potential.

During the subsequent MD simulation, each time step requires evaluating the forces on the sampled potential-energy surface, updating the ionic coordinates and velocities, and determining the electronic energy gap between the two states. For a local machine-learned potential with fixed interaction range, these operations scale approximately linearly with the number of ions, giving $\mathcal{O}(N)$ cost per time step.

For $N_t$ time steps and $N_{\mathrm{traj}}$ independent trajectories, the total MD sampling cost is therefore

\begin{equation}
C_{\mathrm{MD}}
=
\mathcal{O}(N_{\mathrm{traj}}N_tN).
\end{equation}

The autocorrelation function can subsequently be evaluated using FFT-based convolution, requiring $\mathcal{O}(N_{\mathrm{traj}}N_t\log N_t)$ operations approximately \cite{cooleyAlgorithmMachineCalculation1965}. The combined complexity is

\begin{equation}
C_{\mathrm{MD-ACF}}
=
\mathcal{O}\left(
N_{\mathrm{traj}}N_tN
+
N_{\mathrm{traj}}N_t\log N_t
\right),
\end{equation}

which is generally dominated by MD sampling for large systems.

At fixed trajectory length and number of trajectories, the method therefore scales linearly with system size. However, the temporal and statistical sampling requirements also determine the attainable spectral resolution and accuracy. For a fixed MD time step, the frequency resolution associated with a finite correlation window scales inversely with its duration, approximately as $\Delta\omega\propto N_t^{-1}$ \cite{harrisUseWindowsHarmonic1978}. By comparison, the mean spacing of the approximately $3N$ phonon frequencies across a fixed vibrational bandwidth decreases as $N^{-1}$. Resolving spectral features on this scale therefore requires $N_t\propto N$.

Under this additional resolution requirement, the MD sampling cost increases to $\mathcal{O}(N_{\mathrm{traj}}N^2)$. It consequently scales as $\mathcal{O}(N^2)$ for a fixed number of trajectories and as $\mathcal{O}(N^3)$ if the number of independent trajectories also increases proportionally with system size. The latter is a possible additional statistical-sampling requirement, rather than an intrinsic consequence of the MD–ACF formalism. In particular, resolving individual finite-system phonon frequencies is not necessarily required to obtain a converged, smooth optical lineshape. The quadratic-to-cubic scaling estimate is therefore conditional on the chosen spectral-resolution and statistical-convergence criteria.

A major computational advantage of MD–ACF is its comparatively modest memory requirement. At any time step, only the current atomic positions, velocities, forces, and associated model data need to be retained, requiring $\mathcal{O}(N)$ memory. If the electronic energy gap is evaluated during the simulation, storing one scalar value per time step requires an additional $\mathcal{O}(N_t)$ memory. Thus, with sequential trajectory processing, the total working memory scales as

\begin{equation}
M_{\mathrm{MD-ACF}}
=
\mathcal{O}(N+N_t),
\end{equation}

excluding the trained model and reference data. Storing complete atomic trajectories instead would require $\mathcal{O}(NN_t)$ memory per trajectory, but is not necessary for calculating the energy-gap autocorrelation function. Simultaneously processing multiple trajectories increases the working memory approximately in proportion to the number of concurrent simulations.

\subsection{Overall comparison}

The principal computational distinction between the five methods is the stage at which the large-system problem is treated. In direct calculations, the target system size enters the expensive first-principles IFC construction, leading to the representative $\mathcal{O}(N^4)$ scaling. Embedding and interpolation reduce this expense by obtaining the underlying model from a smaller reference supercell. Embedding then constructs an extended real-space vibrational problem, whereas interpolation exploits reciprocal-space decomposition into independent, fixed-size eigenvalue problems.

Machine-learning methods reduce the first-principles cost of obtaining IFCs but generally retain the extended eigenvalue problem, unless additional sparse spectral techniques are employed. MD–ACF avoids diagonalization entirely, replacing it with the temporal and statistical sampling required for converged autocorrelation functions.

These scaling estimates describe the computational structure of the respective approaches rather than universal runtime predictions. Their practical efficiency also depends on numerical prefactors, reference-model accuracy, desired spectral resolution, available parallelization, and the convergence criteria adopted for each method.

\section{Convergence}

\subsection{Locality of the transition-forces \label{sec:nv-force-locality}}

A central assumption of the interpolation procedure introduced in
the main text is that the transition-induced force is sufficiently
localized to be represented within the reference supercell. We therefore investigate
the spatial extent of the mass-weighted transition force $\mathbf{f}^{\textnormal{tr}}_{a}$
obtained from the relaxed ground- and excited-state geometries. Since
the defect is embedded in a periodic supercell, all distances are
evaluated using the minimum-image convention with respect to a force-weighted
cyclic center. For each cell basis-vector direction, this center in normalized (fractional) coordinates is defined
as the circular mean
\begin{equation}
s_{\textnormal{center},\alpha}=\frac{1}{2\pi}\arg\left[\sum_{a}\left|\mathbf{f}^{\textnormal{tr}}_{a}\right|^{2}\exp\left(2\pi i\cdot s_{a,\alpha}\right)\right],\qquad
\mathbf{R}_\textnormal{center}=\begin{pmatrix}
| & | & | \\
\mathbf{h}_1 & \mathbf{h}_2 & \mathbf{h}_3 \\
| & | & |
\end{pmatrix}\mathbf{s}_\textnormal{center},
\end{equation}
where $s_{a,\alpha}$ denotes the normalized (fractional) coordinate of ion $a$
along basis-vector direction $\alpha$, while $\mathbf{R}_\textnormal{center}$ is the Cartesian coordinate vector and $\mathbf{h}_\alpha$ are the basis-vectors. The distance of ion $a$ from this center
is denoted $d_{a}$.

We quantify the localization of the transition-force source using
three complementary measures. The first is the inverse participation
ratio, expressed through the corresponding participation number,
\begin{equation}
N_{\textnormal{part.}}=\frac{1}{IPR}=\frac{\left(\sum_{a}\left|\mathbf{f}^{\textnormal{tr}}_{a}\right|^{2}\right)^{2}}{\sum_{a}\left|\mathbf{f}^{\textnormal{tr}}_{a}\right|^{4}}.
\end{equation}
This quantity gives an effective number of ions contributing significantly
to the transition-force field. The second measure is the cumulative
force weight,
\begin{equation}
C(R)=\frac{\sum_{d_{a}<R}\left|\mathbf{f}^{\textnormal{tr}}_{a}\right|^{2}}{\sum_{a}\left|\mathbf{f}^{\textnormal{tr}}_{a}\right|^{2}},
\end{equation}
or, equivalently, the remaining force weight $1-C(R)$ outside a sphere
of radius $R$. The third measure is the force-weighted root-mean-square
localization radius,
\begin{equation}
R_{\textnormal{RMS}}=\sqrt{\frac{\sum_{a}d^{2}_{a}\left|\mathbf{f}^{\textnormal{tr}}_{a}\right|^{2}}{\sum_{a}\left|\mathbf{f}^{\textnormal{tr}}_{a}\right|^{2}}}.
\end{equation}

\begin{figure}[t]
\centering
\includegraphics{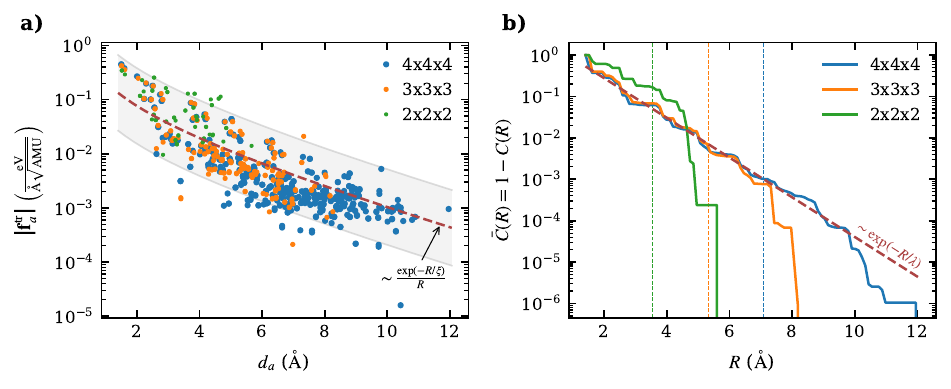}

\caption{\textbf{Localization of the transition-force source for the $\textnormal{NV}^{-}$
optical transition.} \textbf{(a)} Magnitude of the mass-weighted transition
force $\left|\mathbf{f}^{\textnormal{tr}}_{a}\right|$ on each ion
as a function of the minimum-image distance $d_{a}$ from the force-weighted
cyclic center. Results are shown for $2\times2\times2$, $3\times3\times3$,
and $4\times4\times4$ conventional diamond supercells. The exponential trend of the force norm is also shown with a dashed red line and a multiplicative gray band around it of the form $\exp(-R/\xi)/R$, where $\xi=2.968$~Å was taken from the $4\times4\times4$ supercell calculation.  \textbf{(b)}
Residual cumulative force weight $\bar{C}(R)=1-C(R)$, where $C(R)$ is the fraction
of the total squared transition-force weight contained within radius
$R$ from the force-weighted cyclic center. Vertical dashed lines
mark the radius of the largest sphere that can fit inside each supercell. 
An exponential fit is shown for the $4\times4\times4$ supercell data with a dashed red line of 
the form $\exp(-R/\lambda)$, where $\lambda=0.906$~Å was found.}\label{fig:Locality-transition-forces}
\end{figure}

Fig.~\ref{fig:Locality-transition-forces}(a) shows the magnitude
of the transition force on each ion as a function of distance from
the cyclic center for the $2\times2\times2$, $3\times3\times3$,
and $4\times4\times4$ supercells. The force magnitude decreases rapidly
with distance from the defect, consistent with the short-ranged character
expected for a charge-conserving transition between localized defect
states in a gapped host. The scatter at a given distance reflects
the low local symmetry of the relaxed excited-state geometry and the
atomistic structure of the diamond lattice, but the overall envelope
is approximately exponential.

The same behavior is seen more clearly in Fig.~\ref{fig:Locality-transition-forces}(b),
where the residual cumulative weight $1-C(R)$ is plotted on a logarithmic
scale. The decay of $1-C(R)$ demonstrates that the overwhelming majority
of the transition-force weight is contained within the immediate defect
environment. The vertical dashed lines indicate the radius of the
largest sphere that fits inside each supercell. Within the physically
meaningful region before this radius, the $3\times3\times3$ and $4\times4\times4$
results are already very similar near the defect, whereas the $2\times2\times2$
cell shows noticeably larger finite-size effects.

\begin{table}[b]
\centering
\begin{tabular}{|c|c|c|c|}
\hline 
supercell size &
$2\times2\times2$ &
$3\times3\times3$ &
$4\times4\times4$\tabularnewline
\hline 
\hline 
$IPR$ &
0.075 &
0.138 &
0.148\tabularnewline
\hline 
$N_{\textnormal{part.}}$ &
13.294 &
7.248 &
6.748\tabularnewline
\hline 
$R_{\textnormal{RMS}}$ (Å) &
2.595 &
2.154 &
2.099\tabularnewline
\hline 
$\xi$ (Å) &
3.670 &
3.046 &
2.968\tabularnewline
\hline 
\end{tabular}\caption{\textbf{Quantitative localization measures for the transition-force source.}
The table lists the inverse participation ratio, the corresponding
participation number $N_{\textnormal{part.}}$, the force-weighted
RMS localization radius $R_{\textnormal{RMS}}$, and the effective
exponential localization length $\xi$. The localization length is
estimated from $\xi\approx\sqrt{2}R_{\textnormal{RMS}}$, which follows
from an exponential decay model for the residual cumulative force
weight $1-C(R)$.}\label{tab:Locality-measures}
\end{table}

The quantitative localization measures are summarized in Table~\ref{tab:Locality-measures}.
The participation number decreases from $N_{\textnormal{part.}}=13.294$
in the $2\times2\times2$ supercell to $7.248$ and $6.748$ in the
$3\times3\times3$ and $4\times4\times4$ supercells, respectively.
The corresponding force-weighted RMS radius decreases from $2.595$~Å
to $2.154$~Å and $2.099$~Å. This convergence indicates that the
larger apparent spatial extent in the smallest supercell is mainly
a finite-size artifact, while the $4\times4\times4$ supercell gives
a stable representation of the localized transition-force source.

It is useful to relate the cumulative decay to an effective localization
length. In order to avoid unnecessary complexity, assume that the residual force weight follows
\begin{equation}
\bar{C}(R)=1-C(R)\sim\exp\left(-R/\lambda\right).
\end{equation}
Then, in a continuous isotropic three-dimensional approximation,
\begin{equation}
\bar{C}(R)\sim\int^{\infty}_{R}4\pi r^{2}\left|\mathbf{f}^{\textnormal{tr}}(r)\right|^{2}dr.
\end{equation}
This implies an asymptotic force envelope of the form
\begin{equation}
\left|\mathbf{f}^{\textnormal{tr}}(R)\right|\sim\frac{\exp\left(-R/\xi\right)}{R},\quad\xi=2\lambda.
\label{eq:envelopeforces}
\end{equation}

Under the same exponential-tail approximation, the RMS radius satisfies
\begin{equation}
R^{2}_{\textnormal{RMS}}=2\int^{\infty}_{0}R\left[1-C(R)\right]dR=2\lambda^{2},
\end{equation}
and therefore 
\begin{equation}
\xi\approx\sqrt{2}R_{\textnormal{RMS}}.
\end{equation}

Using this estimate, the effective localization length $\xi$ decreases
from $3.670$~Å in the $2\times2\times2$ supercell to $3.046$~Å
and $2.968$~Å in the $3\times3\times3$ and $4\times4\times4$ supercells,
respectively. Since this quantity is derived from $R_{\textnormal{RMS}}$,
it should be regarded as an alternative parametrization of the same
spatial extent rather than as an independent localization measure.
The convergence of the independent localization indicators, namely
the participation number, the force-weighted RMS radius, and the cumulative
force-weight profiles, supports the use of the $4\times4\times4$
transition-force field as an approximation to the isolated-defect
force source. These results support the key locality assumption underlying
the interpolation method: increasing the hypercell size subsequently
refines the sampling of the vibrational continuum rather than changing
the local electron-phonon coupling problem.

\subsection{Nonanalytic force crossover estimate}

As discussed in Sec.~\ref{sec:spatial-extent-of-forces}, the transition-induced force contains a localized
analytic contribution and, in general, an electrostatically mediated
nonanalytic contribution. For a charge-conserving transition with a nonzero
change in permanent electric dipole, the nonanalytic contribution can have an algebraic
far-field decay. Diamond is centrosymmetric, so the piezoelectric acoustic response is absent. The leading dipole-induced long-range acoustic force considered here therefore arises from flexoelectric coupling, which is symmetry-allowed in centrosymmetric insulating crystals \cite{restaBulkTheoryFlexoelectricity2010,hongFirstprinciplesTheoryCalculation2013}. This contribution scales as $r^{-5}$, as shown in Eqn.~(\ref{eq:piezo-flexo-force-tail}) for a dipolar ($l=1$) source with flexoelectric coupling mechanism. We estimate here the distance beyond which
this weak algebraic contribution could become larger than the exponentially
decaying local force.

For the localized, analytic force contribution, we use the envelope introduced in
Sec.~\ref{sec:local-force-contribution} and motivated by the cumulative-force analysis of Sec.~\ref{sec:nv-force-locality},
\begin{equation}
    F^\textnormal{an}(r) \simeq \frac{A}{r}e^{-r/\xi}.
\end{equation}
Figure~\ref{fig:Locality-transition-forces}(a) shows the magnitude of the mass-normalized forces for each ion, $\lvert\mathbf f_a^{\rm
tr}\rvert$, as a function of the distance from the defect center. Around $r^*=3\,\text{\AA}$, a
representative value for the $4\times4\times4$ supercell is
$f^*\simeq 0.05\,\mathrm{eV}/(\text{\AA}\sqrt{\mathrm{AMU}})$, with substantial deviation between atoms at similar distances.
For a carbon atom this representative mass-normalized force corresponds to an absolute force of
$F^*\simeq\sqrt{12\,\mathrm{AMU}}\,f^*\simeq0.17\,\mathrm{eV/\text{\AA}}$ in magnitude.
Taking $\xi\simeq2.968\,\text{\AA}$ from Table~\ref{tab:Locality-measures} gives
\begin{equation}
    A = F^*r^*e^{r^*/\xi} \simeq 1.43\,\mathrm{eV}.
\end{equation}
This is an order-of-magnitude normalization of the local force envelope,
not a fit to the data in Fig.~\ref{fig:Locality-transition-forces}(a). In particular,
the value of $\xi$ is inferred from the RMS localization radius in Sec.~\ref{sec:nv-force-locality}.

To estimate the magnitude of the nonanalytic contribution, we consider the flexoelectric force induced by the electric field of the transition dipole. As discussed in Sec.~\ref{sec:long-range-force-contribution}, varying the flexoelectric interaction energy with respect to the displacement field introduces two spatial derivatives of the dipolar electric field, resulting in a force contribution proportional to the fourth derivatives of $1/r$. In cubic diamond, the isotropic part of this contribution vanishes away from the defect, leaving a long-range force determined by the anisotropy of the longitudinal flexoelectric coefficients, $\delta\mu^D=\mu_{L2}^D-\mu_{L1}^D$. Taking the angular root-mean-square magnitude of this force density and multiplying by the carbon atomic volume $\Omega_C$, we obtain that the acoustic continuum force scales as
\begin{equation}
    F^\textnormal{NA}(r)\sim F^\textnormal{fx}_{(1)}(r) \simeq \frac{B_5}{r^5},
    \qquad
    B_5 \simeq \frac{12\Omega_C\lvert\delta\mu^D\rvert
    \lvert\mathbf p\rvert}{4\pi\varepsilon_0}.
\end{equation}
Here $\mathbf p$ is the difference between the permanent electric dipoles of
the two electronic states, not the optical transition dipole. We take
$\lvert\mathbf p\rvert=0.903\,e\,\text{\AA}$ from Ref.~\cite{galiInitioTheoryNitrogenvacancy2018}
and $\Omega_C\simeq5.55\,\text{\AA}^3$. The fixed-electric-displacement
diamond flexoelectric coefficients reported by Hong and Vanderbilt are
$\mu^D_{L1}=-175.4\,\mathrm{pC/m}$ and
$\mu^D_{L2}=-163.5\,\mathrm{pC/m}$, giving the cubic-anisotropy
combination $\lvert\delta\mu^D\rvert
=\lvert\mu^D_{L2}-\mu^D_{L1}\rvert=11.9\,\mathrm{pC/m}$
\cite{hongFirstprinciplesTheoryCalculation2013}. Within the simplified angular-average and
electrostatic-boundary-condition convention of this estimate, these values
give $B_5\simeq6.43\,\mathrm{eV\,\text{\AA}^4}$.

Equating the two envelopes,
\begin{equation}
    \frac{A}{r_c}e^{-r_c/\xi}=\frac{B_5}{r_c^5},
    \qquad\text{or}\qquad
    r_c^4e^{-r_c/\xi}=\frac{B_5}{A},
\end{equation}
and selecting the outer solution yields $r_c\simeq39\,\text{\AA}$. At this
distance both envelopes are only about
$7\times10^{-8}\,\mathrm{eV/\text{\AA}}$. Using a broader but still
reasonable range of mass-normalized force amplitudes from Fig.~\ref{fig:Locality-transition-forces}(a),
$f_*\simeq0.02$-$0.10\,\mathrm{eV}/(\text{\AA}\sqrt{\mathrm{AMU}})$ at
$3\,\text{\AA}$, shifts the outer crossing to approximately
$35$-$42\,\text{\AA}$. Thus, the crossover is expected
only at a distance where the absolute force is already very small.

However, this comparison is only illustrative. The crossover lies
beyond the range directly sampled in our calculations, and its numerical position
depends on the extrapolated exponential envelope, tensor projection,
screening, and electrostatic boundary conditions. It therefore does not
establish an observed far-field tail, but illustrates why the localized
source can dominate the integrated transition-force weight even if a
nonzero algebraic tail eventually dominates far from the defect.

\section{Assumed baseline for experimental data}

The experimental optical lineshape of the NV$^{-}$ center measured
at 77~K was obtained from Su \textit{et al.} \cite{suLuminescenceLandscapesNitrogenvacancy2019}.
The lineshape was measured on a sample which also contains a non-negligible
concentration NV$^{0}$ centers; see Fig.~3 of Su \textit{et al.} \cite{suLuminescenceLandscapesNitrogenvacancy2019}.
As a result, the measured NV$^{-}$ lineshape contains a contribution
from the tail of the NV$^{0}$ lineshape. Additionally, as noted by Su \textit{et al.} \cite{suLuminescenceLandscapesNitrogenvacancy2019}, the sensitivity of the detector is reduced at longer wavelengths.
Consequently, the experimental lineshape has a non-trivial baseline.

\begin{figure}[bh]
\begin{centering}
\includegraphics[width=10cm]{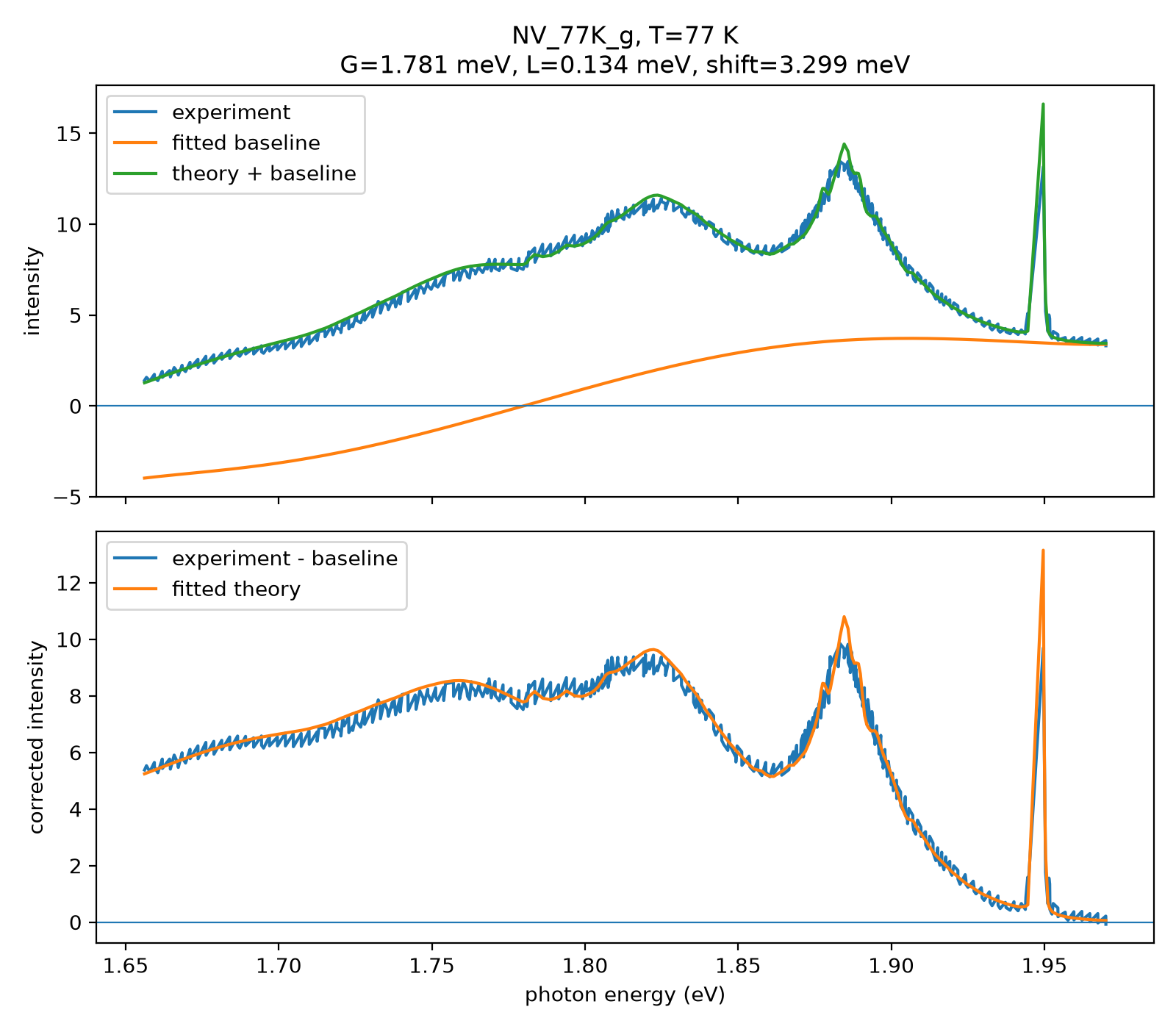}
\par\end{centering}
\caption{\textbf{Assumed baseline for the 77~K experimental lineshape of the NV$^{-}$
center.} The theoretical lineshape and a smooth phenomenological baseline were jointly
fitted to the 77~K experimental spectrum, while the Gaussian (G) and Lorentzian (L) 
FWHM broadening parameters were optimized. The baseline was restricted
to a smooth spline function to avoid overfitting.}\label{fig:baseline}

\end{figure}

To compare the spectra, we fitted the theoretical lineshape calculated 
with the interpolation method introduced here to the experimental data, 
allowing for a smooth phenomenological baseline. The baseline was 
determined by fitting
the theoretical lineshape to the experimental lineshape, while the
additional Gaussian (G) and Lorentzian (L) FWHM broadening parameters
were also optimized. The baseline was restricted to be a sufficiently
smooth function in order to avoid overfitting. The fitted broadening
parameters and baseline are shown in Fig.~\ref{fig:baseline}. The 
fitted baseline becomes negative in part of the spectral range. This 
behavior may be consistent with a wavelength-dependent instrumental 
response, but the baseline is determined phenomenologically from the 
comparison with the theoretical lineshape. Its sign and shape should 
therefore not be interpreted as an independent measurement of detector 
efficiency.

\begin{figure}
\centering{}\includegraphics{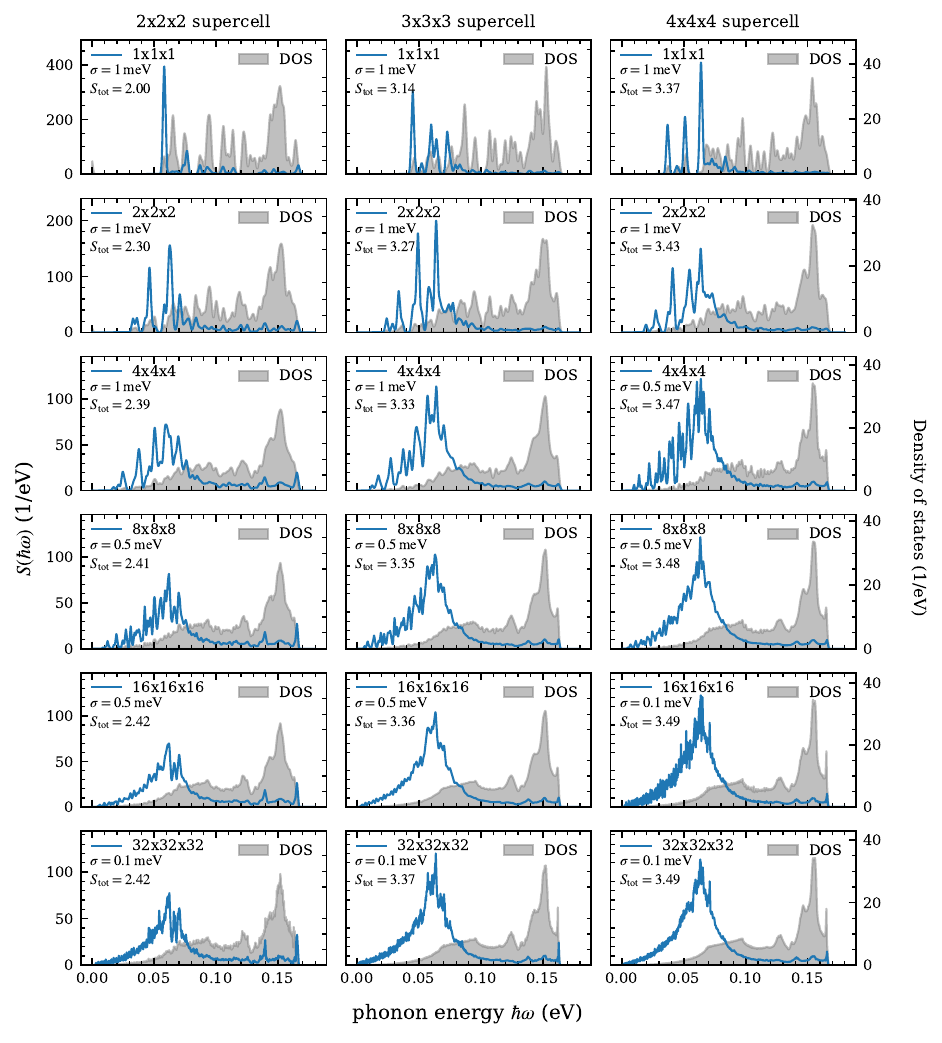}\caption{\textbf{Hypercell convergence of the Huang-Rhys spectral density.} Calculated
Huang-Rhys spectral densities $S(\hbar\omega)$ shown are determined by ground-state
vibrational modes at different defect supercell sizes and
interpolation-grid densities (hypercell sizes). The phonon density of states is shown for
comparison. The dominant coupling band is centered near $63$~meV,
while weaker structures near the optical phonon band edge arise from
van Hove features of the host phonon spectrum. The spectra were obtained
by applying Gaussian broadening to the discrete partial Huang-Rhys factors, where the 
Gaussian standard deviation $\sigma$ is indicated for each case.}\label{fig:Hypercell-convergence-of-HR-spectral-density}
\end{figure}


\end{document}